\documentclass[num-refs]{wiley-article}

\usepackage{siunitx}
\usepackage{natbib}
\usepackage{multirow} 
\usepackage{amssymb}
\usepackage{xcolor,color}
\usepackage{float}
\papertype{Original Article}
\paperfield{Journal Section}

\title{Event-triggered distributed MPC for resilient voltage control of an islanded microgrid}

\author[1]{Pudong Ge}
\author[2]{Boli Chen}
\author[1]{Fei Teng}

\affil[1]{Department of Electrical and Electronic Engineering, Imperial College London, London, SW7 2AZ, UK}
\affil[2]{Department of Electronic and Electrical Engineering, University College London, London, WC1E 6BT, UK}

\corraddress{Fei Teng, Department of Electrical and Electronic Engineering, Imperial College London, London, SW7 2AZ, UK}
\corremail{f.teng@imperial.ac.uk}

\begin{document}

\maketitle

\begin{abstract}
This paper addresses the problem of distributed secondary voltage control of an islanded microgrid (MG) from a cyber-physical perspective. An event-triggered distributed model predictive control (DMPC) scheme is designed to regulate the voltage magnitude of each distributed generators (DGs) in order to achieve a better trade-off between the control performance and communication and computation burdens. By using two novel event triggering conditions that can be easily embedded into the DMPC for the application of MG control, the computation and communication burdens are significantly reduced with negligible compromise of control performance. In addition, to reduce the sensor cost and to eliminate the negative effects of non-linearity, an adaptive non-asymptotic observer is utilized to estimate the internal and output signals of each DG. Thanks to the deadbeat observation property, the observer can be applied periodically to cooperate with the DMPC-based voltage regulator. Finally, the effectiveness of the proposed control method has been tested on a simple configuration with 4 DGs and the modified IEEE-13 test system through several representative scenarios.

\end{abstract}

\section{Introduction}

A microgrid (MG) is a single controllable entity with interconnected loads and distributed energy resources~\cite{ton_u.s._2012,olivares_trends_2014,antoniadou-plytaria_distributed_2017}. Combining these physical plants with indispensable measurement and control loops, MG has been investigated as a typical cyber-physical system (CPS)~\cite{wang_impacts_2019}. A MG can connect and disconnect from the grid to operate in either grid-connected or islanded mode~\cite{ton_u.s._2012,mehrizi-sani_potential-function_2010}. When in the islanded mode, MG control architecture can be divided into three parts: primary control, secondary control and tertiary control~\cite{khayat_secondary_2020,guerrero_hierarchical_2011}. The primary control is implemented locally, whereas the secondary and  tertiary control coordinate the controllable distributed generators (DGs) in the MG to achieve respective control objectives: commonly the objective of the secondary control is to regulate the voltage/frequency to its references and to guarantee the accurate power sharing, while the objective of the tertiary control is to achieve the economic dispatch~\cite{olivares_trends_2014,khayat_secondary_2020,zhang_three-stage_2019}. 

This paper focuses on the secondary control of the MGs. Initial research on this topic investigates the centralized control strategies~\cite{diaz_centralized_2017}, where DGs receive control commands from a center controller. However, due to the fact that the centralized control structure suffers communication delays and requires extensive communication and computation infrastructure, the distributed control strategies, which allow each DG to communicate only with neighboring DGs, have received increasing attention~\cite{cady_distributed_2015,etemadi_decentralized_2012}. In particular, distributed control strategies such as linear feedback control~\cite{bidram_distributed_2013,riverso_plug-and-play_2015,wang_cyber-physical_2019}, finite-time control~\cite{pilloni_robust_2018,abhinav_distributed_2018}, fixed-time control~\cite{zhou_distributed_2020}, have been applied to improve the secondary control in the MG with sparse communication network. Model predictive control (MPC)~\cite{anderson_distributed_2019} has been recently introduced to distributed MG voltage control and demonstrated its superior performance. However, MPC algorithm exacerbates the burden on the online computation and real-time communication due to its prediction mechanism.  Most of existing distributed secondary control methods of the MG~\cite{pilloni_robust_2018,zhou_distributed_2020,zuo_distributed_2016,ge_resilient_2020} are still designed and implemented in a time-triggered fashion, where the sensoring and the controlling are conducted periodically. The time-triggered control could lead to inefficient utilization of computation and communication resources as many data transmissions and calculations are not actually essential to guarantee the control performance.

In this context, the event-triggered control has been proposed for distributed model predictive control (DMPC) to achieve a better trade-off between the control performance and communication and computation burden~\cite{heemels_introduction_2012,lehmann_event-triggered_2013,heemels_networked_2010}. 
The event-triggered mechanism can ease the burden on the communication and even keep resilient against reduced communication resources caused by cyber contingency. So far, several event-triggered secondary control methods have been developed in the MG system with droop-based DGs. However, several problems still remain: (i) the triggering conditions for simultaneously reducing computation and communication have not been fully considered; (ii) the resilience brought by the prediction mechanism of the DMPC to the possible cyber events  has not been fully discussed;  (iii) the existing event-triggered MG control methods~\cite{wang_cyber-physical_2019,zhang_packet_2019} are designed with the assumption that the system state information are fully available, which may not be the case for certain system configuration or requires continuously running of an observer.  


To mitigate the aforementioned problems, a distributed resilient voltage control of an islanded MG is designed based on an event-triggered DMPC and an adaptive non-asymptotic observer. The main contributions of this paper are as follows: 

{\color{black}(i) A novel distributed event-triggered DMPC framework is proposed to restore the voltage for islanded MGs. The proposed DMPC algorithm fully considers the dynamics brought by the DG primary control loop, and improves the control performance owing to its constraint-based optimization. The prediction model of the DMPC also can compensate the effect of communication failure to enhance the system resilience by the update principle of the prediction sequence. In addition, two event triggering conditions which can be easily embedded into the DMPC are designed respectively to reduce computation and communication burden in the cyber layer.} 

(ii) An adaptive non-asymptotic observer is designed to facilitate a cost-effective output-based control framework, which, unlike the Luenberger-like observer~\cite{madonski_survey_2015,ge_extended-state-observer-based_2020}, can operate in an intermittent way due to its deadbeat convergence property; Moreover, the integrated control framework that coordinates the proposed DMPC voltage regulator and the non-asymptotic observer is designed from a timing sequence perspective. 

The remainder of this paper is organized as follows. Section \ref{section_2_problem} is concerned with the cyber-physical modelling of the islanded MG and the corresponding problem formulation. In Section \ref{section_3_Design}, the DMPC with specific event-triggered mechanism and the adaptive non-asymptotic observer are detailed. The corresponding simulation cases are provided in Section \ref{section_4_simulation}, and the conclusions are collected in Section \ref{section_5_conclusion}.   

Primary notations and definitions are given as follows. The set of real numbers is denoted by $\mathbb{R}$. For any vector $\boldsymbol{x}$, $\|\boldsymbol{x}\|$ denotes the Euclidean norm and $\|\boldsymbol{x}\|_{\mathbf{Q}}=\sqrt{\boldsymbol{x}^{T}\mathbf{Q}\boldsymbol{x}}$ stands for $\mathbf{Q}$-weighted norm, where $\mathbf{Q}$ is a matrix with appropriate dimension. The notation $\mathbf{Q}>0$ denotes that $\mathbf{Q}$ is a positive definite matrix. For any set $N$, $|N|$ denotes the number of elements in $N$. For any $n$th order differentiable $y(t)$, $y^{(n)}(t)$ denotes the $n$th order differential value. The notation $\textbf{1}_{n}\in \mathbb{R}^{n}$ denotes a column vector with all elements being ones, i.e., $\textbf{1}_{n}=\left[1,1,\cdots,1\right]^{T}$. The notation $\mathbf{I}_{n}$ denotes the $n$th order identity matrix.


\section{Problem Formulation}\label{section_2_problem}
\begin{figure}[bt]
\centering
\includegraphics[width=3in]{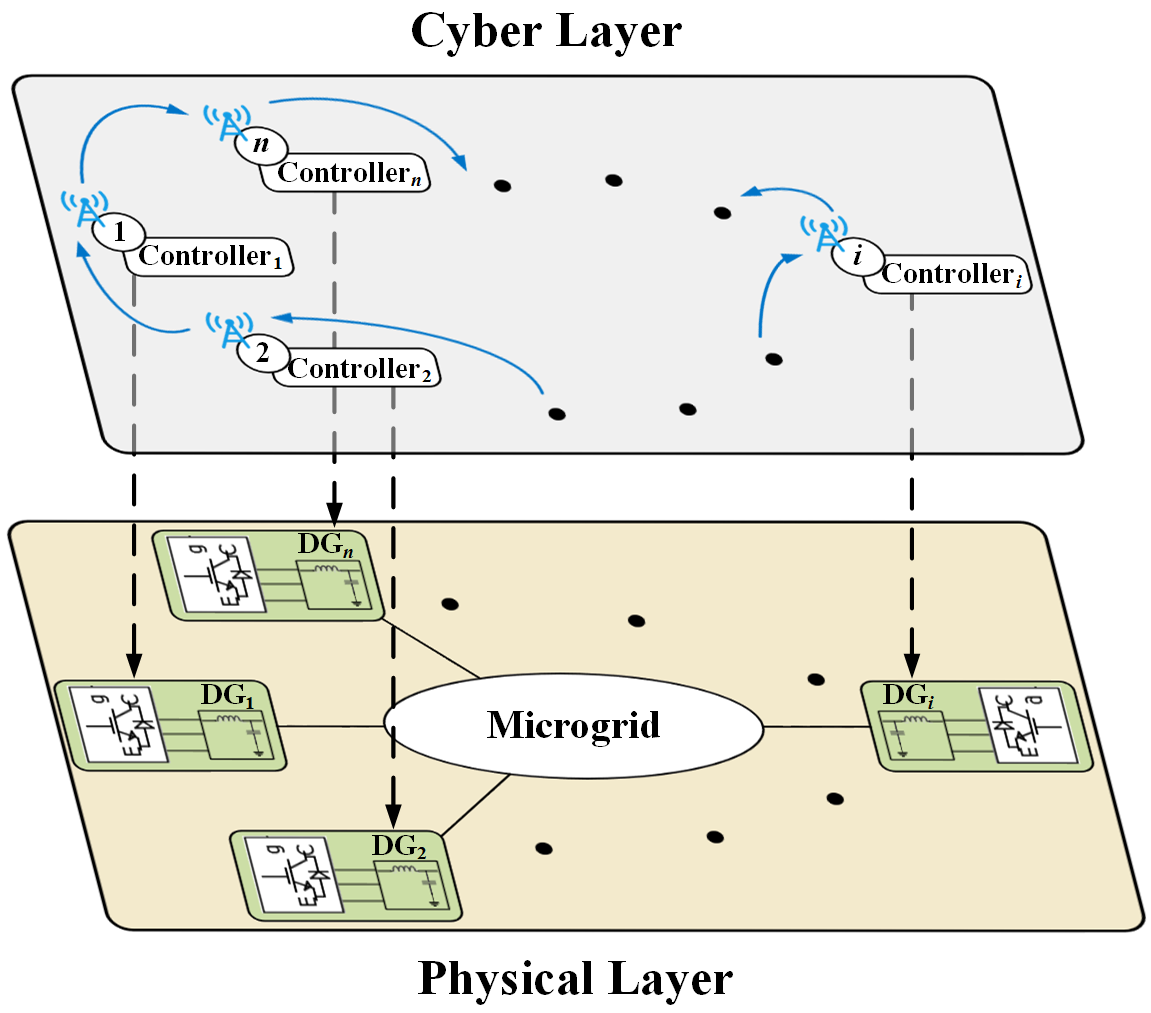}
\caption{Distributed control structure of a cyber-physical coupling MG.}\label{fig_CPS_MG}
\end{figure}

In this section, the model for designing distributed control method of an islanded microgrid is detailed from a cyber-physical coupling system perspective. The physical system contains the electrical topology of the MG and its local controllers, while the cyber layer of the MG can be modeled as a multi-agent system with interconnecting communications, as shown in Figure \ref{fig_CPS_MG}.

\subsection{Physical System}
The MG physically contains multiple DGs that are interconnected through the electrical network. If there is a line between DG $i$ and DG $j$ with the impedance $Z_{ij}=R_{ij}+jX_{ij}$, due to the inductive impedance \cite{wang_cyber-physical_2019,simpson-porco_secondary_2015}, the output active power and reactive power of DG $i$ can be expressed as follows:
\begin{align}
    & P_{i} = P_{iL}+\sum_{j=1}^{N_i}\frac{V_{i}V_{j}}{X_{ij}}\sin{(\theta_{i}-\theta_{j})}\\
    & Q_{i} = Q_{iL}+\sum_{j=1}^{N_i}\left[\frac{V_{i}^{2}}{X_{ij}}-\frac{V_{i}V_{j}}{X_{ij}}\cos{(\theta_{i}-\theta_{j})}\right]
\end{align}
where $P_{iL}$ and $Q_{iL}$ are active and reactive power of the load at bus $i$; and $V_i$ and $\theta_i$ are the bus voltage and the angle at bus $i$. 
{\color{black} In practice, the electrical network connecting DG $i$ and DG $j$ is usually more complicated. However, it is reasonable to model each single MG system by using approximate modelling approaches, where the line impedance is modelled as the equivalent impedance of the network~\cite{floriduz_approximate_2019,dorfler_kron_2013}.
}

Due to the fact that the phase difference $(\theta_{i}-\theta_{j})$ is small~\cite{zhang_distributed_2017}, $\sin{(\theta_{i}-\theta_{j})}\approx(\theta_{i}-\theta_{j})$ and $\cos{(\theta_{i}-\theta_{j})} \approx 1$, which means the active and reactive power can be controlled by the difference of phase angle and voltage magnitude respectively. Thus, the conventional droop control can be obtained:
\begin{align}
    \omega_{i} & = \omega_{ni}-m_{Pi}P_{i} \label{eqn_MG_model_first}\\
    V_{i} & = v_{odi}^{*} = V_{ni}-n_{Qi}Q_{i}
\end{align}
where $\omega_{i},V_{i}$ are the angular frequency and the voltage magnitude provided for the inner control loops. $m_{Pi},n_{Qi}$ are droop coefficients and are selected based on the active and reactive power ratings of each DG \cite{guerrero_hierarchical_2011}. $\omega_{ni},V_{ni}$ are the nominal references of the primary control, which can be generated from the secondary control. It should be noted that each DG is controlled under itself $d$-$q$ (direct-quadrature) axis, which guarantees the voltage magnitude $V_{i}$ is equivalent to the $d$-axis voltage $v_{odi}$, which means $v_{oqi}^{*}=0$. Through the droop control principle, each inverter is controlled with its rotating angular reference.
To model the MG in a uniform frame, a specifically chosen DG is considered as the common reference $\omega_{com}$, and the angular frequency difference of the $i$th DG can be denoted by $\delta_{i}$:
\begin{align}
    \dot{\delta}_{i}=\omega_{i}-\omega_{com}
\end{align}

Combining detailed models in the DG control loops as shown in Figure \ref{fig_DG} (including models of inner loops shown in the APPENDIX), the large-signal dynamic model of the $i$th DG can be detailed as the following multi-input multi-output (MIMO) nonlinear system: 
\begin{align}
     \dot{\boldsymbol{x}}_i = \boldsymbol{f}_i(\boldsymbol{x}_i)+\boldsymbol{g}_i(\boldsymbol{x}_i)\boldsymbol{u}_i+\boldsymbol{k}_i(\boldsymbol{x}_i)\boldsymbol{d}_i(\boldsymbol{x}_j)\label{eq_MG_Model}
\end{align}
with the state vector 
\begin{align*}
    \boldsymbol{x}_i = \left[\delta_i\ P_i\ Q_i\ \phi_{di}\ \phi_{qi}\ \gamma_{di}\ \gamma_{qi}\ i_{ldi}\ i_{lqi}\ v_{odi}\ v_{oqi}\ i_{odi}\ i_{oqi}\right]^T, 
\end{align*}
{\color{black} where the system input is denoted by $\boldsymbol{u}_i=\left[\omega_{ni}\ V_{ni}\right]^T$ with $\omega_{ni}$ and  $V_{ni}$ the input variables for frequency control and voltage control, respectively. $\boldsymbol{d}_i(\boldsymbol{x}_j)=\left[\omega_{com}\ v_{bdi}\ v_{bqi}\right]^T$ represents the interconnection with other DGs, modeled as a disturbance in a single DG system, and $v_{bdi},v_{bqi}$ denote the $d$-$q$-axis voltages at the connection bus in Figure \ref{fig_DG}, which reflects the external disturbance acting on DG $i$.}

\begin{figure}[!bt]
\centering
\includegraphics[width=2.5in]{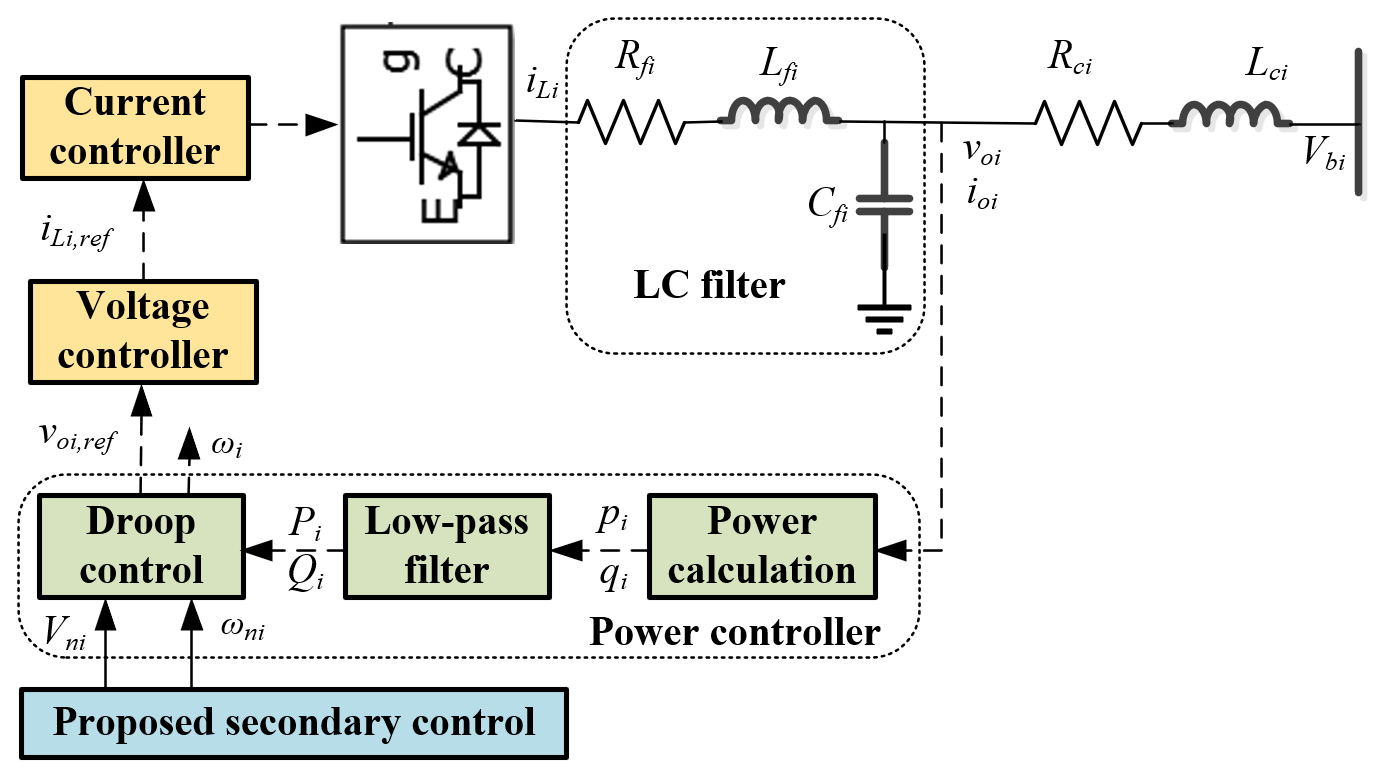}
\caption{Block diagram of the primary control loops in the inverter-based DG.}\label{fig_DG}
\end{figure}

\subsection{Cyber System}
To realize the implementation of the secondary controllers, we assume each DG is equipped with a transceiver for information exchange among sparsely distributed DGs. Thus, as depicted in Figure \ref{fig_CPS_MG}, the communication network in the multi-DG MG can be modelled as a weighted graph $\mathcal{G}_{c}=\{\mathcal{V}_{c},\mathcal{E}_{c}\}$, where $\mathcal{V}_{c}=\{v_{1},v_{2},\dots,v_{N}\}$ is a set of nodes, $\mathcal{E}_{c} \subseteq \mathcal{V}_{c} \times \mathcal{V}_{c}$ is a set of edges, and $N$ is the number of controllable DG nodes. A edge $(v_{j},v_{i})$ means that the $i$th node can receive information from the $j$th node and $v_{j}$ is a neighbour of $v_{i}$. The set of neighbours of node $i$ is described by $N_{i}=\{j:(v_{j},v_{i})\in \mathcal{E}_{c}$. The corresponding adjacency matrix $\mathcal{A}=[a_{ij}]\in \mathbb{R}^{N \times N}$ is denoted by $a_{ii}=0$; $a_{ij}>0$ if $(v_{j},v_{i})\in \mathcal{E}_{c}$, otherwise $c_{ij}=0$. For the graph representing a MG, there exists a virtual leader (reference node), whose adjacency matrix is denoted by $\mathcal{B}={\rm diag}\{b_{i}\}\in \mathbb{R}^{m \times m}$, and the Laplacian matrix $\mathcal{L=D-A+B}$, where $\mathcal{D}={\rm diag}\{\sum_{j\in N_{i}}{a_{ij}}\}$~\cite{zuo_distributed_2016,lewis_secondary_2013}.

The objective of the secondary voltage control designed in the cyber system is to regulate the output voltage magnitude $V_{i}$ of each DG to a unified reference $v_{ref}$ through a leader-following scheme, in the sense that $v_{ref,1}=v_{ref}$ and  $v_{ref,i} = V_{i-1},\forall i >1$. In other words, each DG tracks its neighbors' voltage to achieve the reference tracking. In the cyber layer design, it is meaningful and desirable to limit the computation and communication, especially with the wireless embedded control systems~\cite{heemels_introduction_2012}. From this point of view, this paper proposes an event-triggered control framework, where, as opposed to the conventional control with continuous (or periodic) observation and control of the system, control tasks are executed only when certain conditions are met in order to minimise the computation and communication costs.


\section{Linear DMPC Based Resilient Voltage Control Algorithm Design}\label{section_3_Design}
\begin{figure}[bt]
\centering
\includegraphics[width=3in]{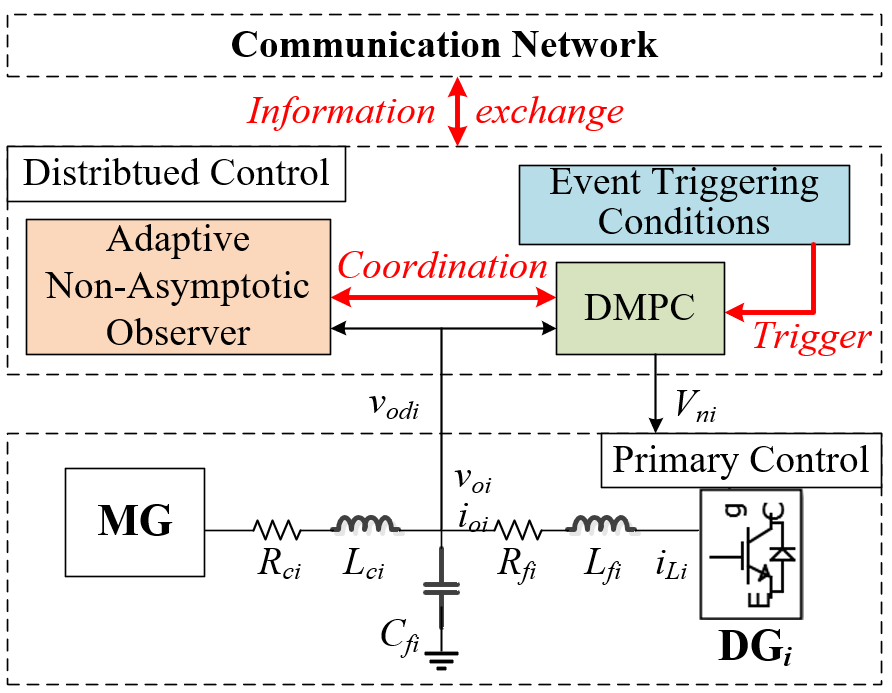}
\caption{Scheme of the DMPC based noise-resilient voltage control.}\label{fig_implementation_scheme}
\end{figure}

The proposed control scheme, as shown in Figure~\ref{fig_implementation_scheme}, is mainly comprised of three parts: distributed model predictive control (DMPC) based voltage regulator, event triggering mechanism design and adaptive non-asymptotic observer. The voltage regulator is designed based on the DMPC framework, where the event-triggered mechanism can be easily embedded to alleviate the computation burden. In addition, the information exchange among agents is also governed by the event-triggered scheme in order to reduce communication cost. 
Finally, to reduce sensor cost, an adaptive non-asymptotic observer is utilized for the reconstruction of internal and output signals. Owing to its fast convergence property, the observer can be operated in an intermittent way, and consequently, it can be integrated into the overall event-triggered control framework.
\subsection{DMPC-Based Voltage Restoration}
The system model \eqref{eq_MG_Model} is a MIMO nonlinear system, but when voltage control is considered, instead of using such a sophisticated model, feedback linearization~\cite{bidram_distributed_2013} is utilized to simplify the model into a linearized form:
\begin{align}
\left\{
    \begin{aligned}
        \dot{y}_{i,1} &= \dot{v}_{odi} = y_{i,2}\\
        \dot{y}_{i,2} &= \ddot{v}_{odi} = f_{i}(\boldsymbol{x}_{i})+g_{i}u_i\\
        y_{i,o} &= y_{i,1} = v_{odi}
    \end{aligned}
    \right.\label{eq_Linear}
\end{align}
\begin{align*}
   & \begin{aligned}
        f_{i}(\boldsymbol{x}_{i})=L_{\boldsymbol{F}_i}^{2}h_i(\boldsymbol{x}_i) = & (-\omega_{i}^{2}-\frac{K_{Pci}K_{Pvi}+1}{C_{fi}L_{fi}}-\frac{1}{C_{fi}L_{ci}})v_{odi} -\frac{\omega_b{K_{Pci}}}{L_{fi}}{v_{oqi}}+\frac{R_{ci}}{C_{fi}L_{ci}}i_{odi}-\frac{2\omega_i}{C_{fi}}i_{oqi}-\frac{R_{fi}+K_{Pci}}{C_{fi}L_{fi}}i_{ldi}\\
        & +\frac{2\omega_i-\omega_b}{C_{fi}}i_{lqi}-\frac{K_{Pci}K_{Pvi}n_{Qi}}{C_{fi}L_{fi}}Q_i+\frac{K_{Pci}K_{Ivi}}{C_{fi}L_{fi}}\phi_{di}+\frac{K_{Ici}}{C_{fi}L_{fi}}\gamma_{di}+\frac{1}{C_{fi}L_{ci}}v_{bdi}
    \end{aligned}\\
    & g_{i}=L_{\boldsymbol{g}_i}L_{\boldsymbol{F}_i}h_i(\boldsymbol{x}_i) = \frac{K_{Pci}K_{Pvi}}{C_{fi}L_{fi}}
\end{align*}
where $f_{i}(\boldsymbol{x}_{i})$ represents the system non-linearity.

Let us define an auxiliary control variable $\xi_{i}= f_{i}(\boldsymbol{x}_{i})+g_{i}u_i$, then $u_i = (g_{i})^{-1}(\xi_i-f_{i}(\boldsymbol{x}_{i}))$ and the dynamic system \eqref{eq_Linear} can be rewritten as 
\begin{align}
    \left\{
    \begin{aligned}
    & \dot{\boldsymbol{y}}_{i}=\mathbf{A}\boldsymbol{y}_{i}+\mathbf{B}\xi_{i} \\
    & y_{i,o}=\mathbf{C}\boldsymbol{y}_{i}
    \end{aligned}
    \right.\label{eqn_MG_voltage_linear}
\end{align}
\begin{align*}
    \boldsymbol{y}_{i}=\left[\begin{array}{c} y_{i,1}\\
    y_{i,2}\end{array}\right], 
    \mathbf{A}=\left[\begin{array}{cc} 0 & 1\\ 0 & 0\end{array}\right],
    \mathbf{B}=\left[\begin{array}{cc} 0 \\ 1\end{array}\right],
    \mathbf{C}=\left[\begin{array}{cc} 1 & 0\end{array}\right]
\end{align*}
The distributed voltage regulation problem is to find appropriate input $\xi_{i}$ to achieve $y_{i,o}\longrightarrow v_{ref,i}$. To implement DMPC, the discrete-time model of \eqref{eqn_MG_voltage_linear} is obtained through Euler discretization: 
\begin{align}
    \left\{
    \begin{aligned}
    & \boldsymbol{y}_{i}(k+1)=\mathbf{A}_{z}\boldsymbol{y}_{i}(k)+\mathbf{B}_{z}\xi_{i}(k) \\
    & y_{i,o}(k)=\mathbf{C}_{z}\boldsymbol{y}_{i}(k)
    \end{aligned}
    \right.\label{eqn_MG_voltage_discrete}
\end{align}
where $\mathbf{A}_{z}=\mathbf{I}+\mathbf{A}T_{s},   \mathbf{B}_{z}=\mathbf{B}T_{s},   \mathbf{C}_{z}=\mathbf{C}$ and $T_{s}$ denotes the sampling time interval.
{\color{black} However, after feedback linearization, 
the dynamics of the discretized system and the real system inevitably differ. An increase in sampling rate will increase the model accuracy whereas computational efficiency degrades. 
To balance the model accuracy and the computational complexity, we design a two-time-scale DMPC model where two time intervals $T_s,T_s^{mpc}$ are defined. $T_s$ denotes the discretization time interval, while $T_s^{mpc}$ denotes the sampling time interval of the DMPC algorithm, and $T_s^{mpc}=rT_s, r\in\mathbb{Z}^+$. Define $h=1,2,\cdots,H$ as the prediction time steps of the DMPC, the full model-based prediction at the time-step $k\ (t_{k+1}-t_k=T_s^{mpc})$ is expressed as
\begin{align}\label{eq:discretization_prediction}
y_{i,o}(k+h_{d}|k) = \mathbf{C}_{z}\mathbf{A}_{z}^{h_{d}}\boldsymbol{y}_{i}(k)+\sum_{i=0}^{h_{d}-1}{\mathbf{C}_{z}\mathbf{A}_{z}^{h_{d}-i-1}\mathbf{B}_{z}{\xi}_{i}(k+i|k)},\qquad h_{d}=1,2,\cdots,Hr
\end{align}
where $h_{d}$ denotes the detailed prediction time steps with length $Hr$ for the discretization model, and the model \eqref{eq:discretization_prediction} also can be expressed in a matrix form:
\begin{align}
\begin{aligned}
\left[\begin{array}{c} y_{i,o}(k+1|k)\\
y_{i,o}(k+2|k)\\ \cdots \\ y_{i,o}(k+Hr|k)\end{array}\right] =\left[\begin{array}{c} \mathbf{C}_{z}\mathbf{A}_{z}\\
\mathbf{C}_{z}\mathbf{A}_{z}^{2}\\ \cdots \\ \mathbf{C}_{z}\mathbf{A}_{z}^{Hr}\end{array}\right]\boldsymbol{y}_{i}(k)
+\left[\begin{array}{cccc} \mathbf{C}_{z}\mathbf{B}_{z} & & &\\     \mathbf{C}_{z}\mathbf{A}_{z}\mathbf{B}_{z} &\mathbf{C}_{z}\mathbf{B}_{z} & &\\ \vdots &\vdots &\ddots & \\ \mathbf{C}_{z}\mathbf{A}_{z}^{Hr-1}\mathbf{B}_{z} &\mathbf{C}_{z}\mathbf{A}_{z}^{Hr-2}\mathbf{B}_{z} &\cdots &\mathbf{C}_{z}\mathbf{B}_{z}\end{array}\right]
\left[\begin{array}{c} \xi_{i}(k|k)\\
\xi_{i}(k+1|k)\\ \cdots \\ \xi_{i}(k+Hr-1|k)\end{array}\right]
\end{aligned}\label{eqn_mpc_prediction_detail}
\end{align}
However, only the prediction at each DMPC time step $k = r,\,2r,\,\ldots$ is required, and therefore the order of the model-based prediction can be reduced and expressed as
\begin{align}
\begin{aligned}
\boldsymbol{Y}_{i,o}(k) &= \left[\begin{array}{c} y_{i,o}(k+r|k)\\y_{i,o}(k+2r|k)\\ \cdots \\ y_{i,o}(k+Hr|k)\end{array}\right] = \left(\mathbf{I}_{H} \otimes \left[\textbf{0}_{1 \times (r-1)}\quad 1\right]\right)\left[\begin{array}{c} y_{i,o}(k+1|k)\\y_{i,o}(k+2|k)\\ \cdots \\ y_{i,o}(k+Hr|k)\end{array}\right] =\left(\mathbf{I}_{H} \otimes \left[\textbf{0}_{1 \times (r-1)}\quad 1\right]\right)\left[\begin{array}{c} \mathbf{C}_{z}\mathbf{A}_{z}\\\mathbf{C}_{z}\mathbf{A}_{z}^{2}\\ \cdots \\ \mathbf{C}_{z}\mathbf{A}_{z}^{Hr}\end{array}\right]\boldsymbol{y}_{i}(k)\\
&+\left(\mathbf{I}_{H} \otimes \left[\textbf{0}_{1 \times (r-1)}\quad 1\right]\right)\left[\begin{array}{cccc} \mathbf{C}_{z}\mathbf{B}_{z} & & &\\     \mathbf{C}_{z}\mathbf{A}_{z}\mathbf{B}_{z} &\mathbf{C}_{z}\mathbf{B}_{z} & &\\ \vdots &\vdots &\ddots & \\ \mathbf{C}_{z}\mathbf{A}_{z}^{Hr-1}\mathbf{B}_{z} &\mathbf{C}_{z}\mathbf{A}_{z}^{Hr-2}\mathbf{B}_{z} &\cdots &\mathbf{C}_{z}\mathbf{B}_{z}\end{array}\right] \left(\mathbf{I}_{H}\otimes\textbf{1}_{r}\right)\left[\begin{array}{c} \xi_{i}(k|k)\\ \xi_{i}(k+1|k)\\ \cdots \\ \xi_{i}(k+H-1|k)\end{array}\right]\\
& = \mathbf{F}_{i}\boldsymbol{y}_{i}(k)+\mathbf{G}_{i}\boldsymbol{\Xi}_{i}(k)
\end{aligned}\label{eqn_mpc_prediction}
\end{align}
where $\boldsymbol{Y}_{i,o}\in\mathbb{R}^{H\times1},\mathbf{F}_{i}\in\mathbb{R}^{H\times2},\mathbf{G}_{i}\in\mathbb{R}^{H\times H}$ and $\boldsymbol{\Xi}_{i}\in\mathbb{R}^{H\times1}$, and more specifically
\begin{align}
	\left[\xi_{i}(k|k)\quad\xi_{i}(k+1|k)\quad\cdots\quad\xi_{i}(k+Hr-1|k)\right]^{T}=\left(\mathbf{I}_{H}\otimes\textbf{1}_{r}\right)\boldsymbol{\Xi}_{i}(k) 
\end{align}
This guarantees the dimension of prediction model \eqref{eqn_mpc_prediction} does not increase, although the full prediction model \eqref{eqn_mpc_prediction_detail} is applied to ensure the prediction accuracy of discretization model. In other words, the prediction sequence $\boldsymbol{Y}_{i,o}(k)$ can be obtained directly from \eqref{eqn_mpc_prediction} instead of \eqref{eqn_mpc_prediction_detail}, and this can save the computation for both prediction and optimization.}
Due to the fact that the proposed DMPC tracking voltage reference by eliminating the difference between local and neighboring DGs' voltage magnitudes, the objective function is designed as follows: 
\begin{align}
    \min \limits_{\boldsymbol{\Xi}_{i}(k)}J_{i}(\boldsymbol{y}_{i}(k),\boldsymbol{\Xi}_{i}(k))= \left\| \frac{1}{|N_{i}|} \sum_{j\in N_{i}}{\boldsymbol{Y}_{i,o}(k)-\boldsymbol{Y}_{j,o}(k)}\right\|_{\mathbf{Q}}^{2} + \left\|\boldsymbol{\Xi}_{i}(k) \right\|_{\mathbf{R}}^{2} \label{eqn_func}
\end{align}
where $|N_{i}|$ denotes the neighbor number of the $i$th DG; the weighting matrix $\mathbf{Q}>0,\mathbf{R}>0$ are designed to balance the tracking performance and the control effort. 
It is noteworthy that when solving the optimization problem, the output of the virtual leader (reference node) is a constant vector $\boldsymbol{Y}_{0,o}(k)=\textbf{1}_{H}v_{ref}$. {\color{black} The synchronization of the voltage signals represents the main target of the application addressed in this paper. For this reason, the weighting factors $\mathbf{Q},\mathbf{R}$ are selected to emphasize the former term in \eqref{eqn_func}.}


{\color{black} Finally, the DMPC framework is completed by the following constraint 
\begin{align}
	0.97p.u.\leq V_i\leq 1.03p.u.
	\label{eq:voltage_constraints}
\end{align}
which restricts the voltage tracking error to $3\%$ so as to enable fast restoration of the voltage to the acceptable range. This constraint can maintain the control performance especially under an exceptional circumstance (e.g., a huge voltage drop or an overvoltage). According to IEEE standard 1547, it is not necessary for the power system to strictly fulfil the constraint \eqref{eq:voltage_constraints} during the operation. However, the tracking error is not permitted to exceed the $3\%$ limit for more than $\bar{T}=0.166$s. In order to meet this requirement, the two sampling interval $T_s$ and $T_s^{mpc}$ calibrated, such that $T_s^{mpc}$ is reasonably smaller than $\bar{T}$ to ensure smooth operation of the system.
The optimization problem \eqref{eqn_func} is solved recursively at each time step $k$ subject to \eqref{eq:voltage_constraints}, and the first control input $\xi_{i}(k|k)$ of the optimal control sequence $\boldsymbol{\Xi}_{i}(k)$ is applied at the $i$th DG.
}

\subsection{Event Triggering Condition Design}
\begin{figure}[bt]
	\centering
	\includegraphics[width=3in]{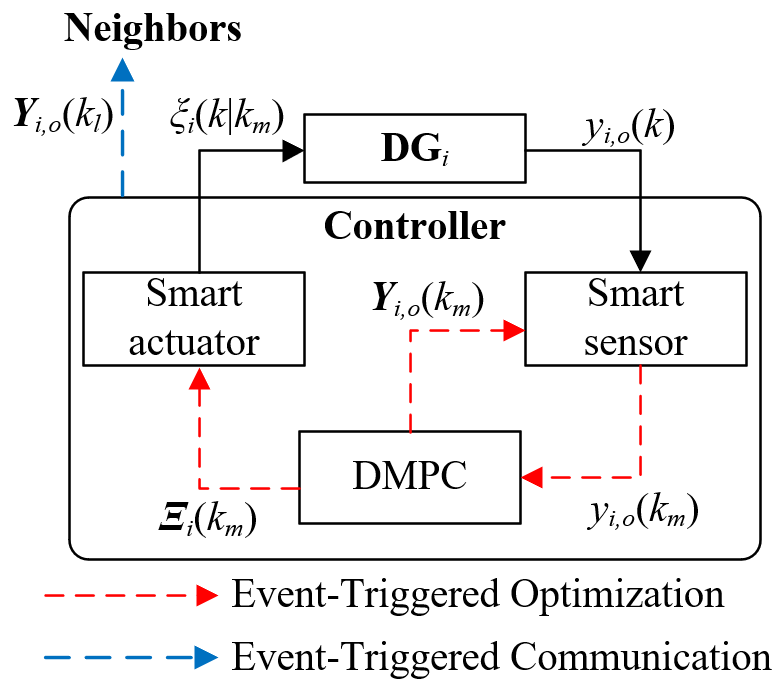}
	\caption{Event-triggered DMPC scheme.}\label{fig_etc_mpc}
\end{figure}

Traditionally, the DMPC-based voltage regulation algorithm relies on the iterative finite-horizon optimization and information exchange among DGs at each time step $k$, which heavily increase the computation and communication burdens. In this connection, an event-triggered scheme is designed and integrated into the DMPC framework to effectively save computation and communication power without sacrificing control performance. The overall scheme of a single DG is shown in Figure~\ref{fig_etc_mpc}. To better demonstrate the event triggering mechanisms, two sets of samples, are defined: $\mathcal{O} = \{k|\Phi(k)\}$ collects the time steps when the DMPC optimization is triggered, where $\Phi(k)$ and $\Psi(k)$ denote the event-trigger rules for optimization and communication, respectively. The design of these rules is introduced next. 

The event-trigger conditions for the DMPC optimization is discussed at first. With the aim of reducing the number of optimization iterations, the DMPC can be made active only when the control performance is not satisfactory. Considering the DMPC is triggered at $k_m$th step ($k_m\in\mathcal{O}$), then for any $k>k_m$ the DMPC is disabled unless 1) the prediction of the system behavior based on the previously calculated control is not reliable any more, or 2) the maximum horizon is reached:
\begin{align}
\Phi(k): \,\,\, \|y_{i,o}(k)-y_{i,o}(k|k_{m})\| \geq e_{opt} \qquad {\rm OR} \qquad k\geq k_{m}+H\label{eqn_tri_opt}
\end{align}
where $e_{opt}>0$ is the user designed threshold for the prediction error. By using this event-triggered optimization mechanism \eqref{eqn_tri_opt}, the stability proof has been discussed in \cite{lehmann_event-triggered_2013} and the tracking error is bounded. Assuming the DMPC is reactivated at $k_m+n$th step with $1 \leq n \leq H$, the control input is not updated by optimization for any steps in between (i.e., $k_m+m,\,1 \leq m < n$). Without loss of generality, the input sequence $\boldsymbol{\Xi}_{i}(k_{m}+m)$ is updated by
\begin{align}
\boldsymbol{\Xi}_{i}(k_{m}+m)=\left[\begin{array}{cccc} \xi_{i}(k_{m}+m|k_{m}) & \cdots & \xi_{i}(k_{m}+H-1|k_{m}) & 0\cdots0 \end{array}\right]^{T},\,1 \leq m < n \leq H \label{eqn_input_update}
\end{align}
and based on \eqref{eqn_input_update} the output predictions are reevaluated by \eqref{eqn_mpc_prediction}.

{\color{black}
On the other hand, to eliminate unnecessary date exchange, the communication between DGs is also regulated by an event-triggered mechanism.
Considering the fact that the communication is not required when the consensus among voltage signals of each DG is achieved, after any communication triggered time step $k_{l}$, the communication is enabled only when the prediction error meets the following condition:
\begin{align}
&\Psi(k):\,\,\,     \|\boldsymbol{Y}_{i,o}(k)-\boldsymbol{Y}_{i,o}(k|k_{l})\|_{\infty}\geq e_{com}\label{eqn_tri_com}, \qquad k> k_{l}\\
&\boldsymbol{Y}_{i,o}(k|k_{l})=\left[\left[\boldsymbol{Y}_{i,o}(k_{l})^{T}\right](k-k_{l}+1)\quad\cdots\quad\left[\boldsymbol{Y}_{i,o}(k_{l})^{T}\right](H)\quad\cdots\quad\left[\boldsymbol{Y}_{i,o}(k_{l})^{T}\right](H)\right]^{T}\label{eq:com_update}
\end{align}
where $\left[*\right](i)$ denotes the $i$th element of the vector. If the communication is not triggered, the neighbours can update the voltage prediction sequence using \eqref{eq:com_update}. This can avoid unnecessary communication if a slight change between two consecutive voltage prediction sequences is captured. As such, if the condition \eqref{eqn_tri_com} is triggered at $k_{l}$th time step ($k_{l}\in\mathcal{C}$), the voltage predictions $\boldsymbol{Y}_{i,o}(k_{l})$ are updated through the communication network. For any $j\in N_{i}$, the differences between the voltage of DG $i$ and the information transmitted to DG $j$ in the DMPC algorithm are bounded by the threshold $e_{com}$ for all $t$.

It should be noted that the voltage prediction remains updated by \eqref{eq:com_update} in presence of communication failure (caused by e.g. packet loss, denial-of-service) between neighbours as the failure interrupts the communication (i.e. communication is not triggered).
In such case, the control terminal value will be the last value in the prediction sequence, which can maintain the performance and enhance the system resilience.
}


Based on the discussion above, the event-triggered DMPC-based voltage regulation algorithm is illustrated in Table \ref{table_etc_mpc}. The impacts of the event triggering thresholds $e_{opt}$ and $e_{com}$ on the system behavior will be numerically investigated in Section~\ref{section_4_simulation} to provide further insights into the selection of the thresholds.

\begin{table}[bt]
\caption{Event-triggered voltage regulation algorithm}\label{table_etc_mpc}
\begin{tabular}{lp{0.9\columnwidth}}
\hline
\headrow
\multicolumn{2}{l}{\textbf{Event-triggered DMPC iterations in time step $k$ for each DG $i$}} \\
\hline
1: & \textbf{given} $k,\boldsymbol{y}_{i}(k),\boldsymbol{Y}_{j,o}(k),j\in N_{i},\boldsymbol{\Xi}_{i}(k-1)$ (update $\boldsymbol{Y}_{j,o}(k)$ from $\boldsymbol{Y}_{j,o}(k-1)$ as \eqref{eq:com_update} if there is no data received): \\
2: & \textbf{if} \eqref{eqn_tri_opt} holds \\
3: & \quad solve \eqref{eqn_func} and \eqref{eq:voltage_constraints} to update the control input sequence $\boldsymbol{\Xi}_{i}(k)$ and the voltage magnitude output sequence $\boldsymbol{Y}_{i,o}(k)$\\
4: & \textbf{else} \\
5: & \quad update $\boldsymbol{\Xi}_{i}(k),\boldsymbol{Y}_{i,o}(k)$ according to \eqref{eqn_input_update} and \eqref{eqn_mpc_prediction} respectively \\
6: & \textbf{end if} \\
7: & apply $\xi_{i}(k|k)$ to DG $i$ \\
8: & \textbf{if} \eqref{eqn_tri_com} holds \\
9: & \quad transmit $\boldsymbol{Y}_{i,o}(k)$ to the neighbours through the communication network\\
10: & \textbf{end if} \\
\hline  
\end{tabular}

\end{table}

\subsection{Finite-time Adaptive Observer Design for Enhancing Noise-Resilience}
The mismatch between the continuous-time system \eqref{eqn_MG_voltage_linear} and the discretized system \eqref{eqn_MG_voltage_discrete} is highly influenced by the non-linearity $f_{i}(\boldsymbol{x}_{i})$ embedded in $\xi_i$ due to the variation of $f_i$ within two samples. As such, the evaluation of the $\boldsymbol{y}_{i}(k+1)$ based on the given control input at $k+1$ may be inaccurate, and in turn, affects the upcoming optimization and prediction. In addition, after generating the auxiliary control variable $\xi_i$, the actual control input $u_i$ is obtained by $u_i = (g_{i})^{-1}(\xi_i-f_{i}(\boldsymbol{x}_{i}))$, where the term $f_{i}(\boldsymbol{x}_{i})$ need to be evaluated and additional sensors may be required to monitor the internal states, such as $v_{odi},\,v_{oqi}$. In fact, to obtain the state $\boldsymbol{y}_{i}$ and the term $f_{i}(\boldsymbol{x}_{i})$, a more cost-effective solution is to use a system observer for reconstructing the real-time state  $\boldsymbol{y}_{i}$ and the time-varying variable $f_{i}(\boldsymbol{x}_{i})$, where the influence of measurement noise can also be highly attenuated~\cite{ge_extended-state-observer-based_2020}. 
{\color{black} The linearized model \eqref{eq_Linear} considering system disturbance for the $i$th DG can be rewritten as:
\begin{align}
    \left\{
    \begin{aligned}
    \dot{y}_{i,1} &= y_{i,2}\\
    \dot{y}_{i,2} &= f'_{i}(\boldsymbol{x}_{i})+g_{i,0}u_i\\
    y_{i,o} &= y_{i,1}= v_{odi}
    \end{aligned}
    \right.\label{eq_extend_model}
\end{align}
\begin{align*}
    &g_i = g_{i,0}+\Delta g_i = L_{\boldsymbol{g}_i}L_{\boldsymbol{F}_i}h_i(\boldsymbol{x}_i)\\
    &f'_{i}(\boldsymbol{x}_{i}) = f_{i}(\boldsymbol{x}_{i})+\Delta g_i u_i
\end{align*}
where $[y_{i,1}\ y_{i,2}]^T$ is the original state vector; $g_{i,0}$ and $\Delta g_i$ denote nominal value and the deviation caused by parameter perturbation of $g_i$, respectively. Moreover, $f'_{i}(\boldsymbol{x}_{i})$ represents the system uncertainty that collects the dynamics of DG inner control loops $f_{i}(\boldsymbol{x}_{i})$, total uncertainties caused by exogenous disturbance, parameter perturbation and the measurement noise.


In the sequel, to streamline the notation, let us consider $\boldsymbol{y}_{i}(t) = \boldsymbol{z}(t)=[z_{0}(t)\quad z_{1}(t)]^{T}$ and $y_{i,o}(t) = y(t)$. Then, the single DG system \eqref{eq_Linear} can be rewritten in the following observer-canonical form:
\begin{align}
    \begin{aligned}
        & \left\{
        \begin{aligned}
            & \dot{\boldsymbol{z}}(t) = \mathbf{A}\boldsymbol{z}(t)+\mathbf{B}u(t)+\mathbf{B}_{w}w(t)\\
            & y(t)=\mathbf{C}\boldsymbol{z}(t)
        \end{aligned}\right.\\
        & \mathbf{A} = \left[\begin{array}{cc}
            a_{1} & 1 \\
            a_{0} & 0
        \end{array}\right],\mathbf{B} = \left[\begin{array}{c}
        b_{1}  \\
        b_{0} 
        \end{array}\right],\mathbf{C} =\left[\begin{array}{cc}
        1 & 0
        \end{array}\right],\mathbf{B}_{w}=\left[\begin{array}{c}
        \alpha_{1}  \\
        \alpha_{0} 
        \end{array}\right] =\left[\begin{array}{c}
        0  \\
        f'(\boldsymbol{x}(t)) 
        \end{array}\right],w(t)=1
    \end{aligned}\label{eqn_DG_observer_form}
\end{align}
with $a_{0}=a_{1}=b_{1}=0,b_{0}=1$.
}

Motivated by a recently proposed deadbeat adaptive observer~\cite{li_distributed_2017}, which offers nearly instantaneous convergence property with high noise immunity, the intermittent (over short time-interval) state and parameter estimation can be enabled to cooperate with the proposed DMPC algorithm. Assuming the short time-interval can guarantee that $f'(\boldsymbol{x}(t))$ can be seen as a constant parameter, we can convert the linear time-varying (LTV) system \eqref{eqn_DG_observer_form} to a linear time-invariant system (LTI) with an unknown parameter $\alpha_{0}=f$. 


To proceed with the analysis, the state-space system \eqref{eqn_DG_observer_form} is expressed as the combination of the input-output derivatives:
\begin{align}
    &y^{(n)}(t)=\sum_{i=0}^{n-1}{a_{i}y^{(i)}(t)}+\sum_{i=0}^{n-1}{b_{i}u^{(i)}(t)}+\sum_{i=0}^{n-1}{\alpha_{i}w^{(i)}(t)}\label{eqn_IOy}\\
    &z_{r}(t) = y^{(r)}(t)-\sum_{j=0}^{r-1}{a_{n-r+j}y^{(j)}(t)}-\sum_{j=0}^{r-1}{b_{n-r+j}u^{(j)}(t)}-\sum_{j=0}^{r-1}{\alpha_{n-r+j}w^{(j)}(t)}
\end{align}
where $n=r=2$ and $\sum_{j=0}^{k}{\{\cdot\}}=0,k<0$. $y^{(n)}(t)$ denotes the $n$th differential value of $y(t)$ and $z_{r}(t)$ denotes the $r$th element of the state in \eqref{eqn_DG_observer_form}.

Let us introduce the Volterra integral operator $V_{K}$ induced by a bivariate function $K(t,\tau)$ to the output and its derivatives:
\begin{align}
    [V_{K}y^{(i)}](t)\triangleq \int_{0}^{t}K(t,\tau)y^{(i)}(\tau)d\tau, \forall i\in \{0,\cdots,n\}\label{eqn_volterra}
\end{align}
where $K(t,\tau)$ is the $n$th order non-asymptotic kernel~\cite{pin_kernel-based_2013} subject to
\begin{equation}\label{eq:non-asymptotic}
K^{(i)}(t,0)=0,\forall i\in \{0,\cdots,n\}    
\end{equation}
After some algebra, we get:
\begin{align}
    [V_{K}y^{(i)}](t)=\sum_{j=0}^{i-1}{(-1)^{i-j-1}y^{(j)}(t)K^{(i-j-1)}(t,t)}+(-1)^{i}[V_{K^{(i)}}y](t) \label{eqn_non_asymptotic}
\end{align}
which can be obtained by applying the integral by parts and \eqref{eq:non-asymptotic}.
If $i=1$,
\begin{align}
    [V_{K^{(1)}}y](t)=y(t)K(t,t)-[V_{K}y^{(1)}](t)\label{eqn_non_asymptotic_1}
\end{align}
Replacing $y(t)$ with $y^{(n-1)}(t)$, \eqref{eqn_non_asymptotic_1} becomes 
\begin{align*}
    [V_{K^{(1)}}y^{(n-1)}](t)=y^{(n-1)}(t)K(t,t)-[V_{K}y^{(n)}](t)
\end{align*}
which can be further expanded by substituting \eqref{eqn_IOy} 
\begin{align}
    \begin{aligned}
        (-1)^{n-1}[V_{K^{(n)}}y](t)=&-\sum_{j=0}^{n-2}{(-1)^{n-2-j}y^{(j)}(t)K^{(n-j-1)}(t,t)}+y^{(n-1)}(t)K(t,t)\\
        &-\sum_{i=0}^{n-1}a_{i}[V_{K}y^{(i)}](t)-\sum_{i=0}^{n-1}b_{i}[V_{K}u^{(i)}](t)-\sum_{i=0}^{n-1}\alpha_{i}[V_{K}w^{(i)}](t)
    \end{aligned}\label{eqn_non_asymptotic_2}
\end{align}
Substituting \eqref{eqn_non_asymptotic} and its same forms with $u(t),w(t)$ into \eqref{eqn_non_asymptotic_2}, we obtain
\begin{align}
    \begin{aligned}
        (-1)^{n-1}[V_{K^{(n)}}y](t)+\sum_{i=0}^{n-1}{(-1)^{i}a_{i}[V_{K^{(i)}}y](t)}+\sum_{i=0}^{n-1}{(-1)^{i}b_{i}[V_{K^{(i)}}u](t)}\\
        =-\sum_{i=0}^{n-1}{(-1)^{i}\alpha_{i}([V_{K^{(i)}}w](t)}+\sum_{r=0}^{n-1}{(-1)^{n-r-1}K^{(n-r-1)}(t,t)z_{r}(t)}
    \end{aligned}\label{eqn_non_asymptotic_3}
\end{align}
where the state variables $z_{r}(t)$ and the unknown parameters $\alpha_{i}$ appear explicitly, and can be obtained by the casual filtering of the signals $y(t),u(t)$.


Considering the specific parameters of \eqref{eqn_DG_observer_form}, the following expression can be inferred from \eqref{eqn_non_asymptotic_3}:
\begin{align}
    \begin{aligned}
        (-1)[V_{K^{(2)}}y](t)+[V_{K}u](t)       =f[V_{K}w](t)+(-1)K^{(1)}(t,t)z_{0}(t)+K(t,t)z_{1}(t)
    \end{aligned}\label{eqn_non_asymptotic_4}
\end{align}
To estimate the state and unknown parameter, let us define
\begin{align}
    &\lambda(t)\triangleq (-1)[V_{K^{(2)}}y](t)+[V_{K}u](t)\label{eqn_lamda}\\
    &\boldsymbol{\gamma}(t)\triangleq\left[[V_{K}w](t),(-1)K^{(1)}(t,t),K(t,t)\right]\label{eqn_gamma}
\end{align}
Then, \eqref{eqn_non_asymptotic_4} can be rewritten as
\begin{align}
    \lambda(t)=\boldsymbol{\gamma}(t)\left[\begin{array}{c}
         f \\
         \boldsymbol{z}(t)
    \end{array}\right]\label{eqn_dbo}
\end{align}
To find the estimates of $\left[\begin{array}{cc}
         f & \boldsymbol{z}(t)
    \end{array}\right]^{T}$ (of dimension 3), we can apply three
    different non-asymptotic kernel functions to augment \eqref{eqn_dbo} into three linearly independent equations
\begin{align}
    \boldsymbol{\Lambda}(t)=\boldsymbol{\Gamma}(t)\left[\begin{array}{c}
         f  \\
         \boldsymbol{z}(t)
    \end{array}\right]\label{eqn_dbo_1}
\end{align}
where $\boldsymbol{\Lambda}(t)=\left[\lambda_{0}(t),\lambda_{1}(t),\lambda_{2}(t)\right]^{T}$ and $\boldsymbol{\Gamma}(t)=\left[\boldsymbol{\gamma}^{T}_{0}(t),\boldsymbol{\gamma}^{T}_{1}(t),\boldsymbol{\gamma}^{T}_{2}(t)\right]^{T}$, and $\lambda_{h}(t),\boldsymbol{\gamma}_{h}(t),h\in\{0,1,2\}$ are \eqref{eqn_lamda} and \eqref{eqn_gamma} induced with the kernel functions respectively. The three kernel functions are designed as follows~\cite{pin_kernel-based_2013}:
\begin{align}
    K_{h}(t,\tau)=e^{-\omega_{h}(t-\tau)}(1-e^{-\varpi\tau})^{2}, h\in\{0,1,2\}\label{eqn_kernel_func}
\end{align}
which meets the non-asymptotic condition \eqref{eq:non-asymptotic}.  
Finally, the estimates are obtained by:
\begin{align}
    \left[\begin{array}{c}
         \hat{f}  \\
         \hat{\boldsymbol{z}}(t)
    \end{array}\right]=\boldsymbol{\Gamma}^{-1}(t)\boldsymbol{\Lambda}(t),\forall t_{\epsilon}<t<t_{\epsilon}+\Delta t \label{eqn_dbo_voltage}
\end{align}
where $t_{\epsilon}$ is the observer initialization time to guarantee the invertibility of $\boldsymbol{\Gamma}(t)$ ($\boldsymbol{\Gamma}(0) =0$) and $\Delta t$ is the active time of the observer. 
{\color{black} The observer ensures finite and instantaneous convergence of the state estimates to the true state with high level of noise immunity.}
The detailed discussion about the robustness of the observer is show in~\cite{pin_robust_2019}.

\begin{figure}[bt]
\centering
\includegraphics[width=5in]{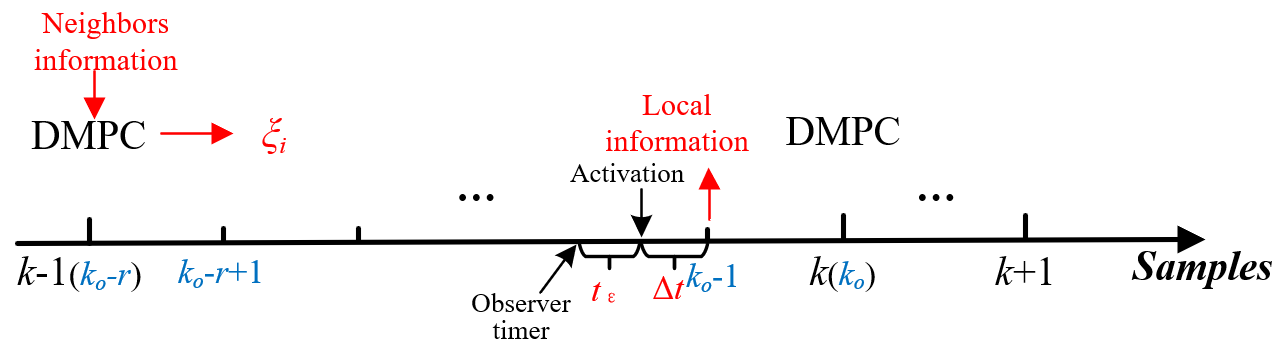}
\caption{Time-sequence cooperation between the event-triggered DMPC and the non-asymptotic observer.}\label{fig_etmpc_dbo}
\end{figure}

{\color{black}
The non-asymptotic observer is sampled at $T_s$ and it cooperates with the event-triggered DMPC voltage regulation in a periodical manner, 
as shown in Figure \ref{fig_etmpc_dbo}. To ensure the estimates, $\hat{f}_{i}$ and $\hat{\boldsymbol{y}}_{i}$, available for the voltage regulator at each DMPC sampling instant. The observer is always enabled $\Delta t+t_{\epsilon}$ seconds ahead of an MPC step. For example, assuming the time at the $k$-th MPC step is $t_{k_o}$, the proposed observer is enabled at $t_{k_o}-\Delta t-t_{\epsilon}$, and after the holding time $t_{\epsilon}$ the estimates start updating. Both estimates $\hat{f}_{i}(t_{k_o})$ and $\hat{\boldsymbol{y}}_{i}(t_{k_o})$ are fed to the voltage regulator at $t_{k_o}$ when the observer is disabled.}

\section{Simulation Results}\label{section_4_simulation}
In this section, the proposed event-triggered control method is tested on a simple MG configuration with 4 DGs  and on the modified IEEE-13 test system.
\begin{figure}[htb]
\centering
\includegraphics[width=4in]{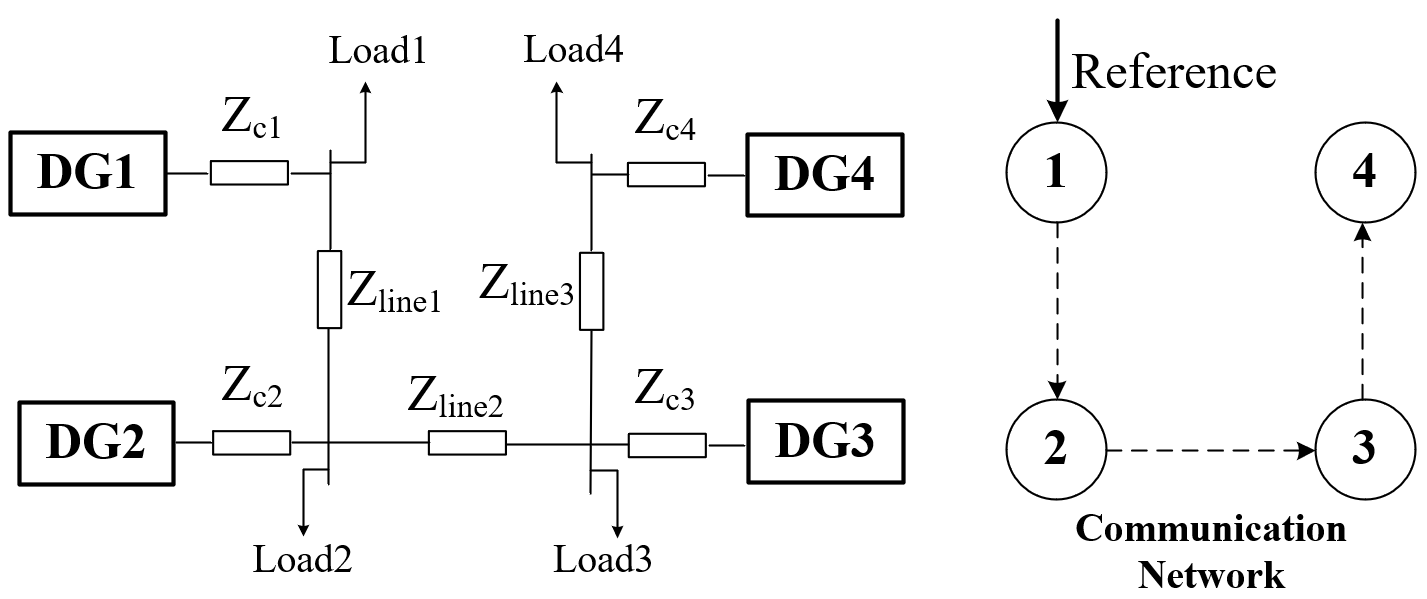}
\caption{Diagram of the tested 4-bus MG system.}\label{fig_case_MG}
\end{figure}
\begin{figure}[!htb]
\centering
\includegraphics[width=3.5in]{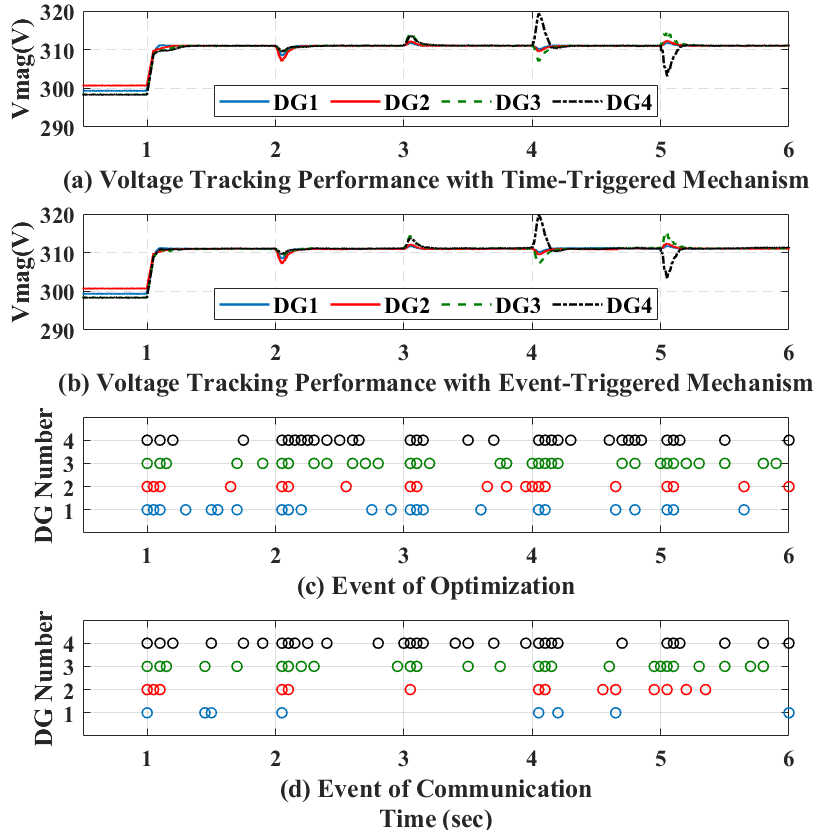}
\caption{Voltage control performance by using event-triggered mechanism: (a) voltage tracking performance with time-triggered mechanism; (b) voltage tracking performance with event-triggered mechanism; (c) event-triggered time of DMPC optimization; (d) event-triggered time of neighbouring communication.}\label{fig_case1_1}
\end{figure}
\subsection{Case 1: 4-DG MG system}
The single line diagram of the 4-DG MG and its communication topology is shown in Figure \ref{fig_case_MG}. The parameters of the tested MG system and the proposed controllers is shown in Table \ref{table_para_MG1} (see Appendix). The simulation test involves a few representative scenarios by which the effectiveness of the proposed methodology can be reflected.

\subsubsection{Scenario 1: Load Change and Plug-and-Play Capability Test}
In this Scenario, the control performance of the proposed control is illustrated under load change and DG's plug-and-play operation: in the beginning, Load2 is disconnect from the system and only primary control is applied; at $t=1\mathrm{s}$, the proposed secondary control is activated; Load2 and half of Load3 are connected and disconnected at $t=2\mathrm{s}$ and $t=3\mathrm{s}$ respectively, and DG4 is disconnected and re-connected at $t=4\mathrm{s}$ and $t=5\mathrm{s}$ respectively. The performance of voltage tracking is shown in Figure~\ref{fig_case1_1} and the reductions of computation and communication are detailed in Table~\ref{table_case1_1}. 

\begin{table}[!htb]
\centering
\caption{Computation and communication reductions by using event-triggered mechanism}\label{table_case1_1}
\begin{tabular}{lccccc}
\hline
\headrow
& \thead{DG1} & \thead{DG2} & \thead{DG3} & \thead{DG4} & \thead{Average} \\
\hline
Computation Reduction & 77\% & 80\% & 68\% & 66\% & 72.75\% \\
Communication Reduction & 92\% & 86\% & 74\% & 69\% & 80.25\% \\
\hline  
\end{tabular}

\end{table}

\begin{figure}[!htb]
\centering
\includegraphics[width=5.5in]{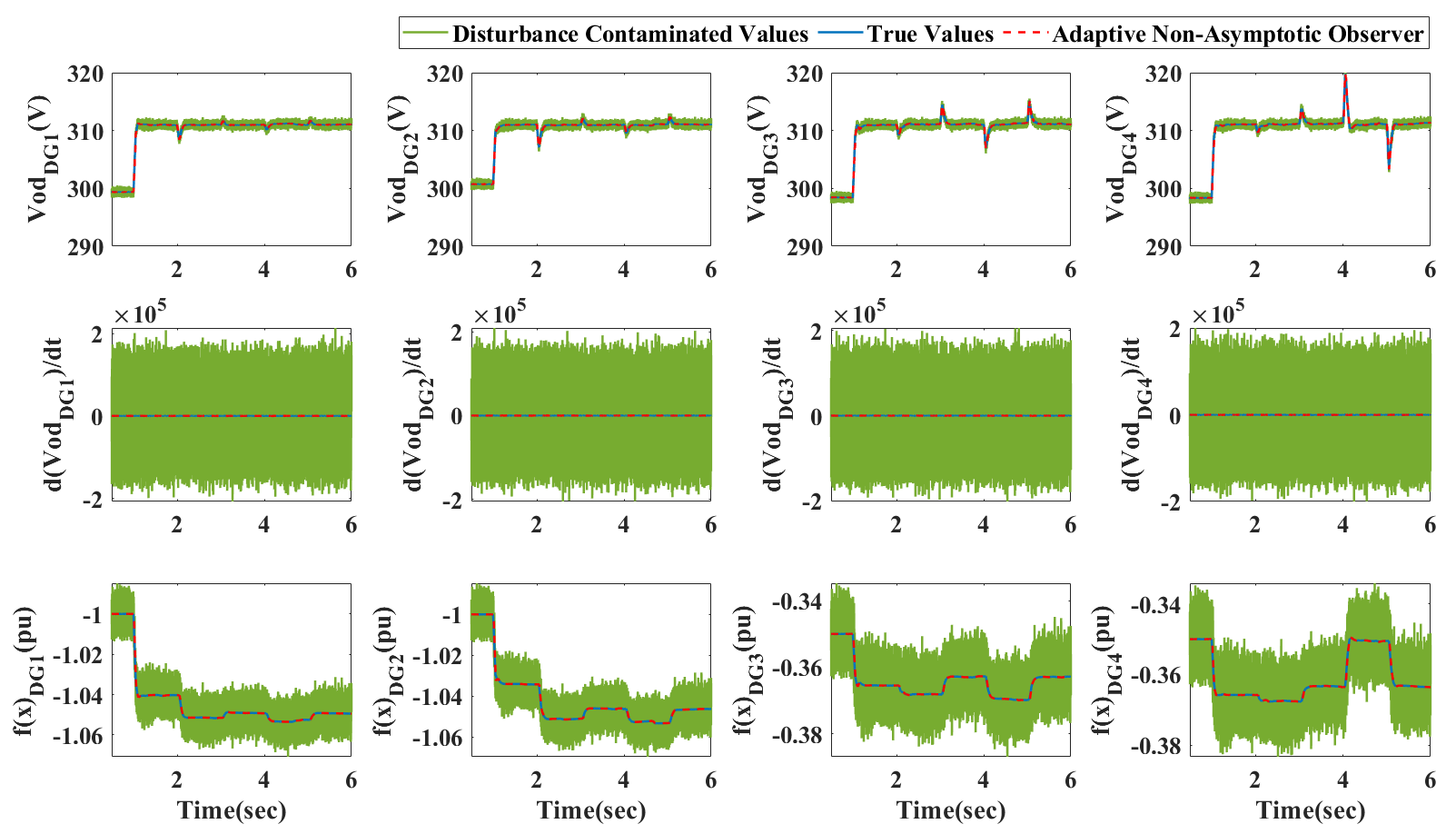}
\caption{Non-asymptotic observer performance (base value of $f{(x)}$: $7.35\times10^{9}$).}\label{fig_case1_2}
\end{figure}

\begin{figure}[!htb]
\centering
\includegraphics[width=5.5in]{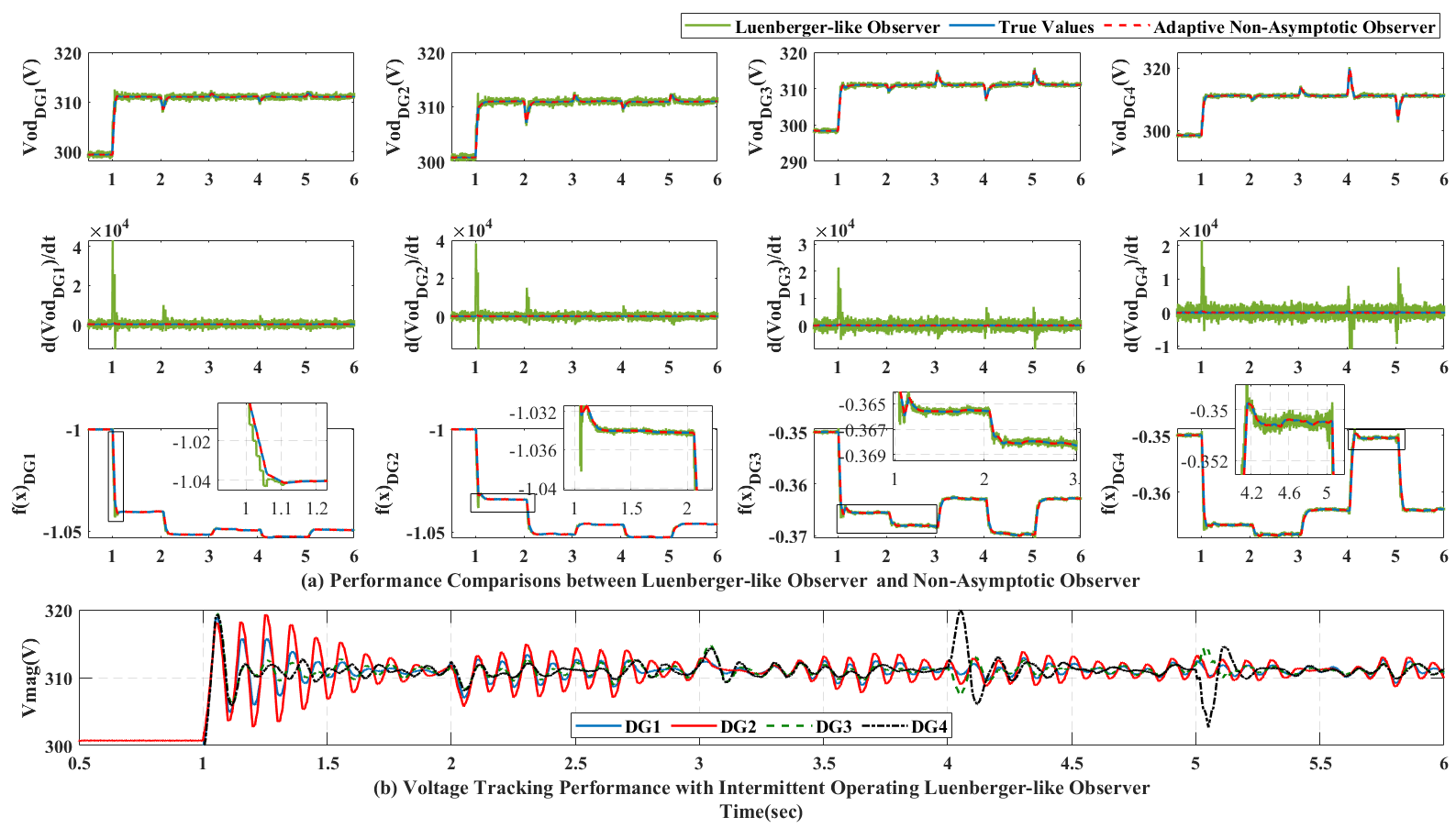}
\caption{Voltage control performance with intermittent operating Luenberger-like observer.}\label{fig_case1_3}
\end{figure}
By using the event-triggered mechanism, the sacrifice of control performance is limited, whereas the computation and communication are both considerably reduced. By employing the proposed non-asymptotic observer, the negative effects of the disturbance can be eliminated, as shown in Figure \ref{fig_case1_2}. The performance of the proposed observer is emphasized by the comparisons among true values, observed values and disturbance contaminated values that are obtained from indirect measurement in the noisy environment. Compared to the previous Luenberger-like extended state observer~\cite{ge_extended-state-observer-based_2020}, the proposed non-asymptotic observer benefits from its intermittent operating characteristic. The performance comparisons between intermittent operating Luenberger-like observer and the proposed non-asymptotic observer is shown in Figure \ref{fig_case1_3}, where we can see that Luenberger-like Observer cannot estimate the state precisely when the system responses to the physical events. If the Luenberger-like extended state observer is working intermittently as the proposed non-asymptotic observer, the voltage tracking performance will degrade as Figure \ref{fig_case1_3}(b). 

{\color{black} To further illustrate  the resilient performance of the DMPC-based algorithm, an extreme condition with a dramatic voltage drop has been simulated. At t=2s, DG4 is disconnected from the MG while the loads increase, thus the DG output voltage may drop to the unacceptable sections (out of the constraint \eqref{eq:voltage_constraints}). Figure \ref{fig_DMPC_constraints} compares the control performance between DMPC-based and PI-based algorithms. When using DMPC-based algorithm, the voltage magnitudes are restored into the constraints faster due to the voltage constraints. However, the PI-based algorithm, as a linear control method, cannot handle such a voltage drop efficiently.
	
\begin{figure}[!htb]
	\centering
	\includegraphics[width=5.5in]{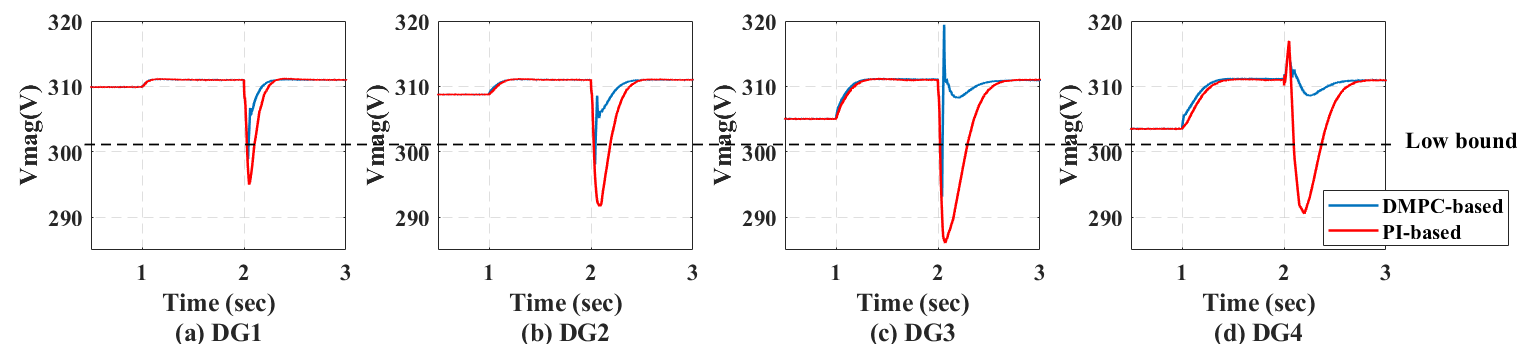}
	\caption{Voltage control performance under the extreme condition.}\label{fig_DMPC_constraints}
\end{figure}
}

\subsubsection{Scenario 2: Control Performance with Different Event Triggering Thresholds}
The control performance of proposed event-triggered mechanism may be influenced by the selection of thresholds for both computation and communication event generators. Therefore, in Scenario 2, case studies as Scenario 1 are carried out with different triggering thresholds.
\begin{figure}[!htb]
\centering
\includegraphics[width=5in]{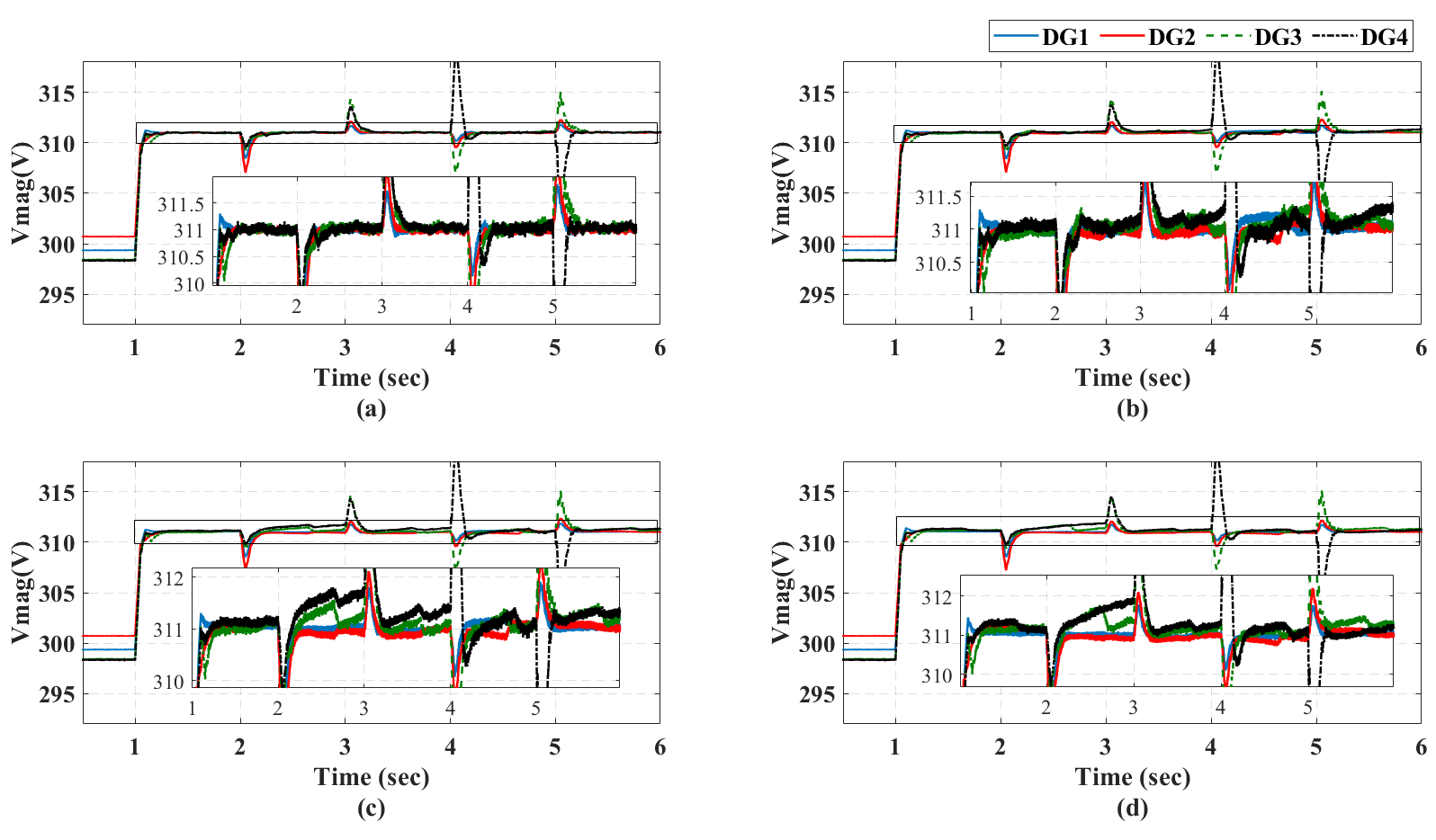}
\caption{Event-triggered condition with fixed $e_{com}$ (${e}_{com}=0.1$) but different thresholds $e_{opt}$: (a) $e_{opt}=0.05$; (b) $e_{opt}=0.1$; (c) $e_{opt}=0.15$; (d) $e_{opt}=0.2$.}\label{fig_case2_1}
\end{figure}

\begin{figure}[!htb]
\centering
\includegraphics[width=5in]{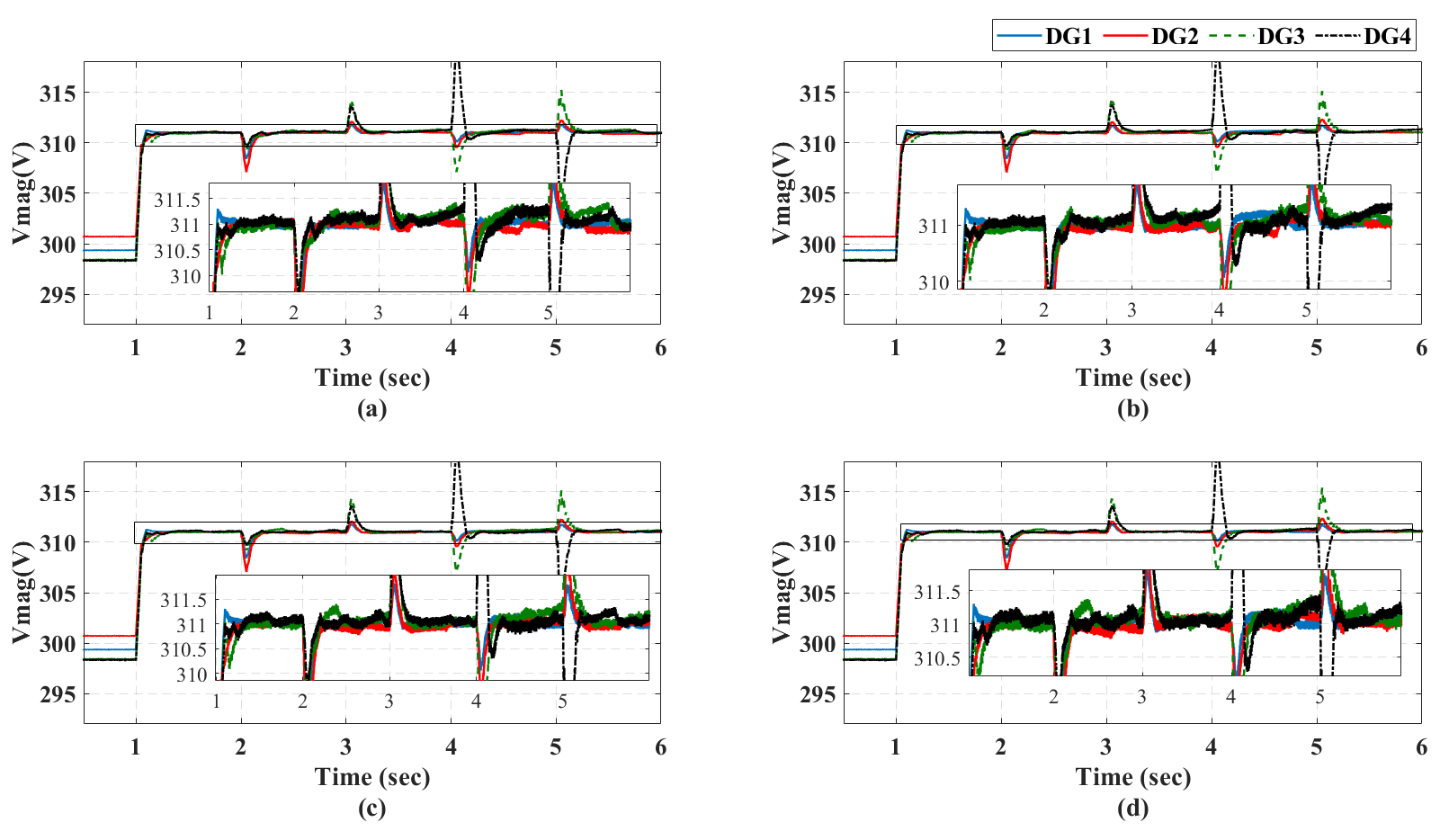}
\caption{{\color{black}Event-triggered condition with fixed $e_{opt}$ ($e_{opt}=0.1$) but different thresholds $e_{com}$: (a) $e_{com}=0.05$; (b) $e_{com}=0.1$;(c) $e_{com}=0.15$; (d) ${e}_{com}=0.2$.}}\label{fig_case2_2}
\end{figure}

\begin{table}[!htb]
\centering
\caption{Computation and communication reductions with fixed $e_{com}$ (${e}_{com}=0.1$) but different thresholds $e_{opt}$.}\label{table_case2_1}
\begin{tabular}{lcccccc}
\hline
\headrow
& \thead{$e_{opt}$} & \thead{DG1} & \thead{DG2} & \thead{DG3} & \thead{DG4} & \thead{Average} \\
\hline
\hiderowcolors
\multirow{4}{*}{Computation Reduction} & 0.05 & 24\% & 24\% & 34\% & 34\% & 29\% \\
& 0.1 & 77\% & 80\% & 68\% & 66\% & 72.75\% \\
& 0.15 & 83\% & 84\% & 83\% & 81\% & 82.75\% \\
& 0.2 & 87\% & 84\% & 85\% & 83\% & 84.75\% \\
\hline
\multirow{4}{*}{Communication Reduction} & 0.05 & 95\% & 89\% & 84\% & 84\% & 88\% \\
& 0.1 & 92\% & 86\% & 74\% & 69\% & 80.25\% \\
& 0.15 & 88\% & 77\% & 65\% & 64\% & 73.5\% \\
& 0.2 & 86\% & 78\% & 64\% & 65\% & 73.25\% \\
\hline  
\end{tabular}
\end{table} 
The control performance with fixed $e_{com}$ (${e}_{com}=0.1$) but different thresholds $e_{opt}$ is detailed in Figure \ref{fig_case2_1} and Table \ref{table_case2_1}. As $e_{opt}$ increases, the optimization computation of each DG controller decreases largely, but from Figure~\ref{fig_case2_1}, we can also see the control performance will clearly degrade when $e_{opt}=0.2$ and $e_{opt}=0.3$. Thus, the selection of $e_{opt}$ is a trade-off between the tracking performance and the computation reduction. The control performance with fixed $e_{opt}$ ($e_{opt}=0.1$) but different thresholds $e_{com}$ is detailed in Figure~\ref{fig_case2_2} and Table~\ref{table_case2_2}.
{\color{black} As $e_{com}$ increases, the communication among DGs is reduced with the gradually degraded control performance.
}

\begin{table}[!htb]
\centering
\caption{Computation and communication reductions with fixed $e_{opt}$ ($e_{opt}=0.1$) but different thresholds $e_{com}$.}\label{table_case2_2}
\begin{tabular}{lcccccc}
\hline
\headrow
& \thead{$e_{com}$} & \thead{DG1} & \thead{DG2} & \thead{DG3} & \thead{DG4} & \thead{Average} \\
\hline
\hiderowcolors
\multirow{4}{*}{Computation Reduction} & 0.05 & 77\% & 79\% & 69\% & 71\% & 74\% \\
& 0.1 & 77\% & 80\% & 68\% & 66\% & 72.75\% \\
& 0.15 & 79\% & 80\% & 73\% & 64\% & 74\% \\
& 0.2 & 78\% & 75\% & 69\% & 71\% & 73.25\% \\
\hline
\multirow{4}{*}{Communication Reduction} & 0.05 & 67\% & 42\% & 42\% & 41\% & 48\% \\
& 0.1 & 92\% & 86\% & 74\% & 69\% & 80.25\% \\
& 0.15 & 96\% & 91\% & 81\% & 80\% & 87\% \\
& 0.2 & 96\% & 93\% & 86\% & 88\% & 90.75\% \\
\hline  
\end{tabular}
\end{table}

{\color{black} \subsubsection{Scenario 3: Effects of Information Update Frequency and Prediction Horizons}

In Scenario 3, the effects of information update frequency and prediction horizon on the control performance are investigated, shown in Figure \ref{fig_case301}. Figure \ref{fig_case301}(a) illustrates the voltage response for different update intervals ($T_s^{mpc}=0.05\mathrm{s},0.1\mathrm{s},0.15\mathrm{s}$). Although the voltage control performance degrades slightly on convergence time as the update interval increases, the computation and communication ($T_s^{mpc}=0.15$s) are reduced significantly by 32.1\% and 68.4\% respectively compared to that of $T_s^{mpc}=0.05$s.
The effect of the prediction horizon is shown on the voltage control performance as prediction horizon decreases. It can be noted that the declining prediction horizon leads to degrading control performance and at the same time higher computation and communication rates (increasing by 70.6\% and 81.0\% respectively as horizon decreases from 10 to 2).

\begin{figure}[!htb]
	\centering
	\includegraphics[width=4in]{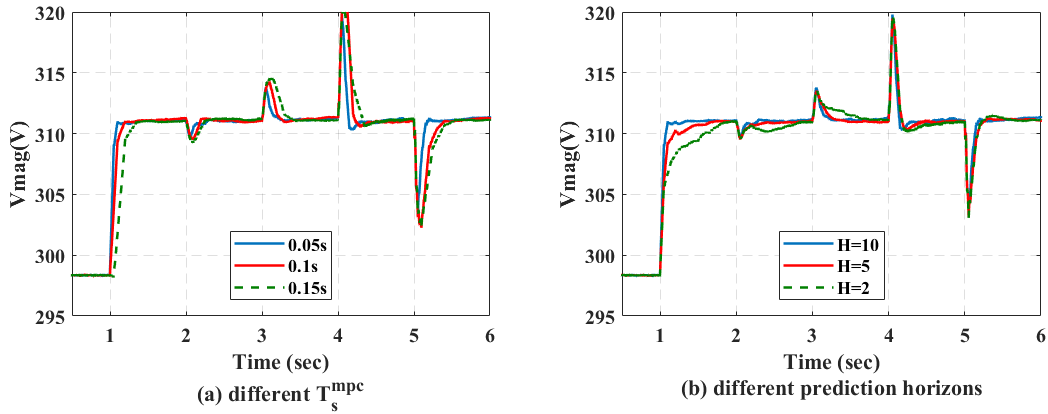}
	\caption{Effects of information update frequency and prediction horizon.}\label{fig_case301}
\end{figure}
}

\subsubsection{Scenario 4: Communication Topology Change}
{\color{black} 
\begin{figure}[!htb]
\centering
\includegraphics[width=4in]{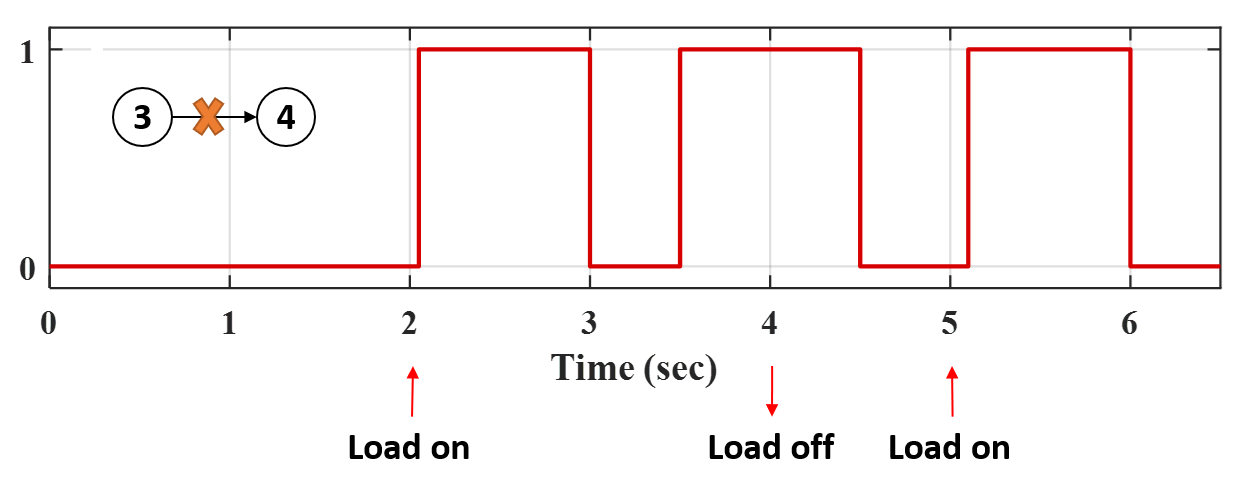}
\caption{Physical and cyber events of the 4-DG MG system: value "1" represents that the communication channel between DG3 and DG4 is unavailable; the load change occurs at 2s, 4s and 5s respectively.}\label{fig_case1_change_new}
\end{figure}

\begin{figure}[!htb]
\centering
\includegraphics[width=3in]{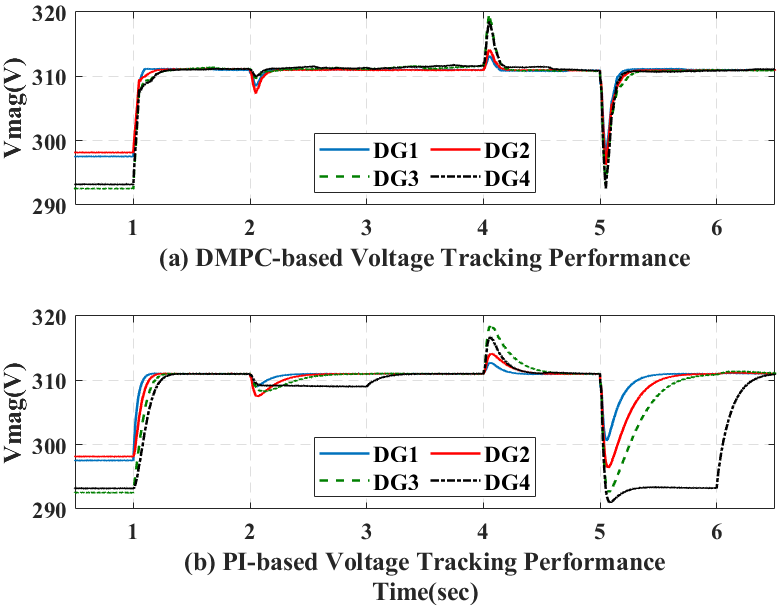}
\caption{Voltage control performance with cyber and physical events.}\label{fig_case3_1_new}
\end{figure}

In Scenario 4, we consider communication interruptions which may occur in the distributed operation, and the physical and cyber events is shown in Figure \ref{fig_case1_change_new}. In the cyber layer, the communication change mimics the failure and recovery of cyber links. In practice, the recovery of communication links takes a finite period of time depending on the numbers of attacked nodes and broken communication links~\cite{ding_distributed_2019}. In this scenario, from 2s to 6s, several failure and recovery events occur. The corresponding control performance is shown in Figure~\ref{fig_case3_1_new}. The voltage tracking performance is maintained during the whole event, although DG4 does not always have the neighbouring information over the time period $2\mathrm{s}<t<6\mathrm{s}$. This is due to the prediction mechanism in the DMPC algorithm, under which DG4 can update the neighbouring information according to the information received before the communication failure occurs. In other words, the prediction model in the event-triggered DMPC helps maintain the control performance in this extreme condition, which enhances the operational resilience. However, the PI-based control can only use the last received information before the communication failure, so it could lead to the tracking error if the system has not entered into the steady state at the time instant when the communication failure occurs. Due to that communication failure can be caused by many practical reasons such as denial of service, actual faults, it is reasonable that there exists load change during the communication failure, thus the proposed DMPC-based control will show better resilience in practice.
}

\subsection{Case 2: Modified IEEE-13 bus system}
\begin{figure}[!htb]
\centering
\includegraphics[width=4in]{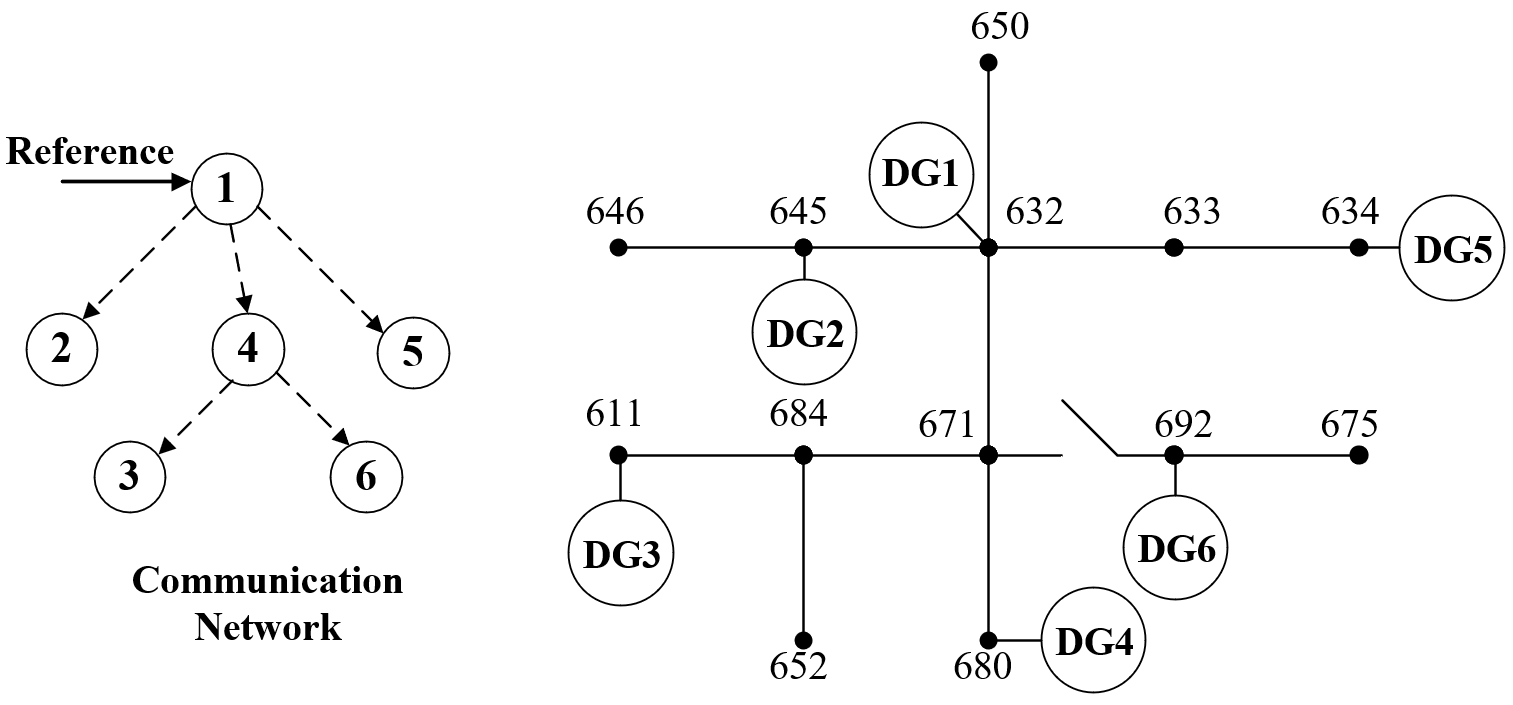}
\caption{Diagram of modified IEEE-13 bus MG system}\label{fig_case_MG_2}
\end{figure}

A real MG system is utilized to further test the effectiveness of the proposed method. The electrical and communication topology of the modified IEEE-13 bus test system~\cite{morstyn_distributed_2015} is shown in Figure~\ref{fig_case_MG_2}, where there is a breaker between node 671 and 692. 
{\color{black} The parameters of 6 DGs are the same as those shown in TABLE \ref{table_para_MG1} (DG5 is the same as DG4, DG6 is the same as DG1). The controller parameters remain the same as well. Due to the fact that this subsection focuses on the scalability and especially the resilience against potential system reconfiguration. The event triggering thresholds are set to $e_{opt}=0.1,e_{com}=0.1$ by following a similar tuning process elaborated in subsection 4.1.2.}

\subsubsection{Scenario 1: Scalability Test}
\begin{figure}[!htb]
\centering
\includegraphics[width=4in]{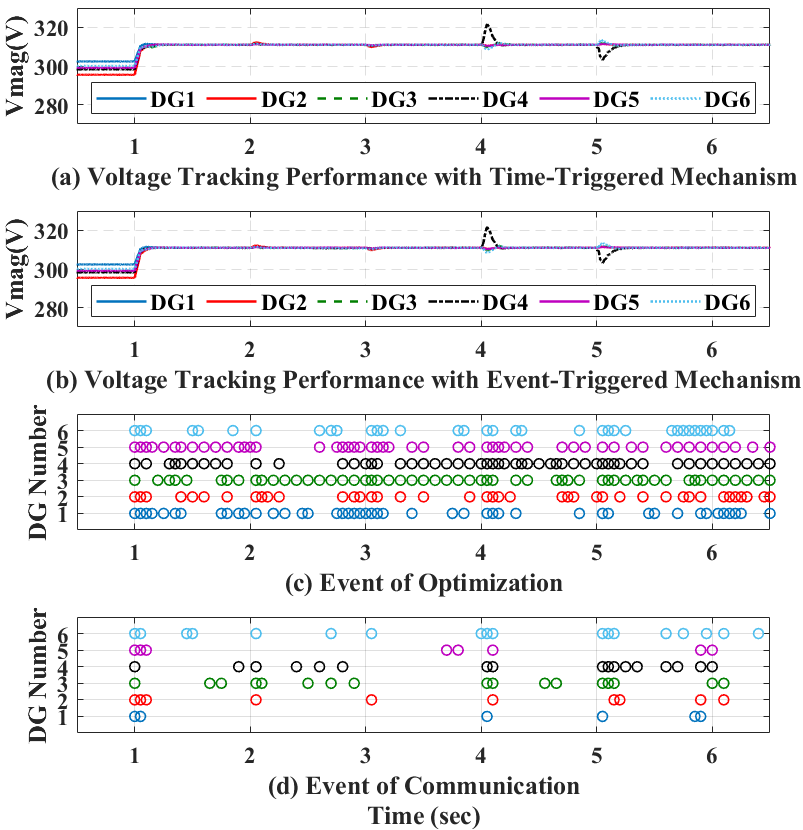}
\caption{Voltage control performance of modified IEEE-13 bus MG system: (a) voltage tracking performance with time-triggered mechanism; (b) voltage tracking performance with event-triggered mechanism; (c) event-triggered time of DMPC optimization; (d) event-triggered time of neighbouring communication.}\label{fig_case4_1}
\end{figure}

In this Scenario, the breaker between nodes 671 and 692 is always switched on, and the scalability of the proposed control is illustrated by load change and DG's plug-and-play scenario: loads at bus 645 and bus 675 are decreased and increased at $t=2\mathrm{s},3\mathrm{s}$ respectively; and DG4 is disconnected and re-connected at $t=4\mathrm{s}$ and $t=5\mathrm{s}$ respectively. The voltage tracking performance is shown in Figure~\ref{fig_case4_1} and the average reductions of computation and communication are 57.42\% and 88.48\%.

\subsubsection{Scenario 2: Resilience Illustration with System Reconfiguration}

\begin{figure}[!htb]
\centering
\includegraphics[width=5.5in]{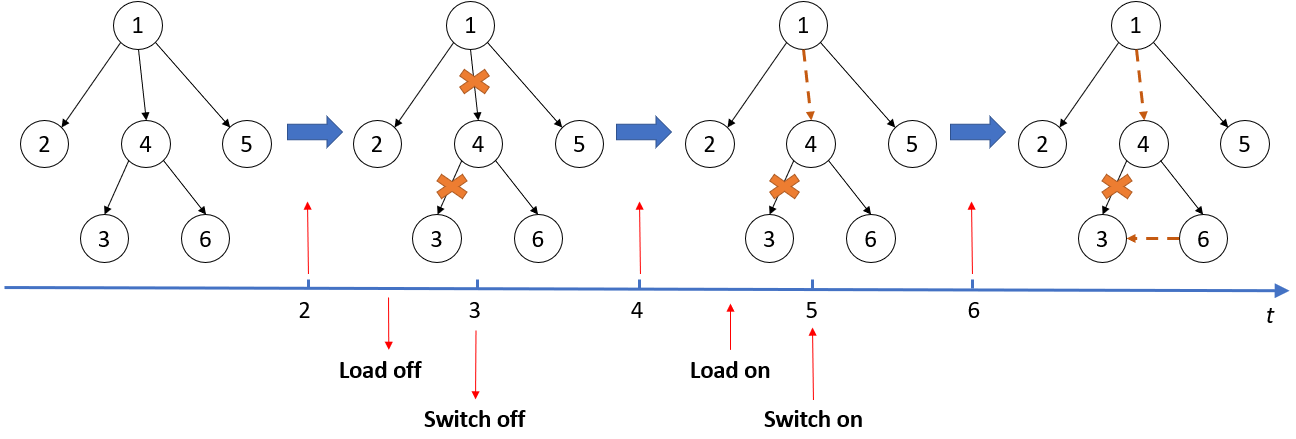}
\caption{Physical and cyber events of modified IEEE-13 bus MG system}\label{fig_case2_change}
\end{figure}

{\color{black}
To evaluate the resilience of the proposed voltage regulation method when the system reconfiguration occurs on both physical and cyber layers, we design the physical and cyber events (including breaker switched off and on) as shown in Figure~\ref{fig_case2_change}. The corresponding control performance is shown in Figure \ref{fig_case5_1}. Although there are both physical and cyber events, similar to the subsection 4.1.4, the voltage tracking performance is guaranteed by using event-triggered DMPC method, and the average reductions of computation and communication are 63.94\% and 88.03\%. The oscillations at $t=5\mathrm{s}$ are incurred by the re-synchronization after the break is switched on.

\begin{figure}[!htb]
\centering
\includegraphics[width=4in]{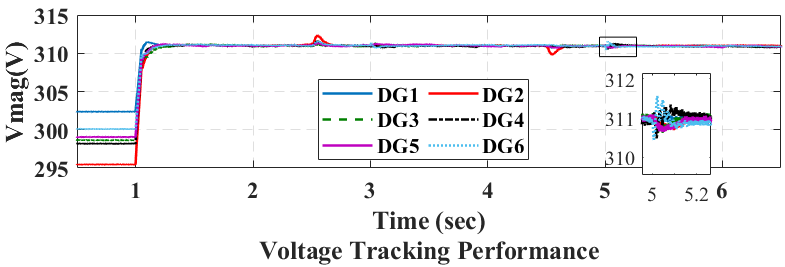}
\caption{Voltage control performance with system reconfiguration in the modified IEEE-13 bus system}\label{fig_case5_1}
\end{figure}
}

\section{Conclusion}\label{section_5_conclusion}
This paper proposes an event-triggered DMPC for secondary voltage control scheme in a cyber-physical coupled MG system, which explicitly considers the model non-linearity and the system noise-resilience. In the control design, based on the event-triggered DMPC, two thresholds are designed to trigger the local DMPC computation and neighboring communications among DGs. To facilitate a cost-effective and noise-resilient control, an adaptive observer that features the non-asymptotic convergence characteristic is utilized, and this designed adaptive non-asymptotic observer can be coordinated with the DMPC voltage regulator in a timing sequence. Finally, the effectiveness of the proposed control method is verified on a 4-DG MG system and the modified IEEE-13 system.

\section*{Appendix}\label{section:appendix}
\subsection*{Dynamic models of DG inner loops}
As shown in Figure \ref{fig_DG}, the instantaneous active and reactive powers are generated through a low-pass filter with the cutoff frequency $\omega_{ci}\ll \omega_{i}$:
\begin{align}
    \dot{P}_{i}=-\omega_{ci}P_{i}+\omega_{ci}(v_{odi}i_{odi}+v_{oqi}i_{oqi})\\
    \dot{Q}_{i}=-\omega_{ci}Q_{i}+\omega_{ci}(v_{oqi}i_{odi}-v_{odi}i_{oqi})
\end{align}
where $v_{odi},v_{oqi}$ and $i_{odi},i_{oqi}$ are $d$-$q$ voltage and current of the $i$th DG output respectively.
Apart from the droop control, the inner control loops (the voltage control loop and the current control loop) are modelled as:

\begin{align}
    \left\{
    \begin{aligned}
        & {{{\dot{\phi }}}_{di}}=v_{odi}^{*}-{{v}_{odi}} \\ 
        & {{{\dot{\phi }}}_{qi}}=v_{oqi}^{*}-{{v}_{oqi}} \\ 
        & i_{ldi}^{*}={{F}_{i}}{{i}_{odi}}-{{\omega }_{b}}{{C}_{fi}}{{v}_{oqi}}+{{K}_{PVi}}(v_{odi}^{*}-{{v}_{odi}})+{{K}_{IVi}}{{\phi }_{di}} \\ 
        & i_{lqi}^{*}={{F}_{i}}{{i}_{oqi}}\text{+}{{\omega }_{b}}{{C}_{fi}}{{v}_{odi}}+{{K}_{PVi}}(v_{oqi}^{*}-{{v}_{oqi}})+{{K}_{IVi}}{{\phi }_{qi}}\\
        & {{{\dot{\gamma }}}_{di}}=i_{ldi}^{*}-{{i}_{ldi}} \\ 
        & {{{\dot{\gamma }}}_{qi}}=i_{lqi}^{*}-{{i}_{lqi}} \\ 
        & \dot{v}_{ldi}^{*}=-{{\omega }_{b}}{{L}_{fi}}{{i}_{lqi}}+{{K}_{PCi}}(i_{ldi}^{*}-{{i}_{ldi}})+{{K}_{ICi}}{{\gamma }_{di}} \\ 
        & \dot{v}_{lqi}^{*}={{\omega }_{b}}{{L}_{fi}}{{i}_{ldi}}+{{K}_{PCi}}(i_{lqi}^{*}-{{i}_{lqi}})+{{K}_{ICi}}{{\gamma }_{qi}}
    \end{aligned}\right.
\end{align}
where $\phi_{di},\phi_{qi}$ and $\gamma_{di},\gamma_{qi}$ are auxiliary variables for the voltage controller and the current controller respectively; $K_{PVi}, K_{IVi}$ and $K_{PCi}, K_{ICi}$ are P-I control parameters for the voltage controller and the current controller; $\omega_{b}$ represents the rated frequency of the MG; $F_{i}$ is the parameter for $d$-$q$ frame compensation.
The dynamics of the LC filter and the output impedance also can be expressed as
\begin{align}
    \left\{
    \begin{aligned}
        & {{{\dot{i}}}_{ldi}}=-\frac{{{R}_{fi}}}{{{L}_{fi}}}{{i}_{ldi}}+{{\omega }_{i}}{{i}_{lqi}}+\frac{1}{{{L}_{fi}}}{{v}_{idi}}-\frac{1}{{{L}_{fi}}}{{v}_{odi}} \\ 
        & {{{\dot{i}}}_{lqi}}=-\frac{{{R}_{fi}}}{{{L}_{fi}}}{{i}_{lqi}}-{{\omega }_{i}}{{i}_{ldi}}+\frac{1}{{{L}_{fi}}}{{v}_{iqi}}-\frac{1}{{{L}_{fi}}}{{v}_{oqi}} \\ 
        & {{{\dot{v}}}_{odi}}={{\omega }_{i}}{{v}_{oqi}}+\frac{1}{{{C}_{fi}}}{{i}_{ldi}}-\frac{1}{{{C}_{fi}}}{{i}_{odi}} \\ 
        & {{{\dot{v}}}_{oqi}}=-{{\omega }_{i}}{{v}_{odi}}+\frac{1}{{{C}_{fi}}}{{i}_{lqi}}-\frac{1}{{{C}_{fi}}}{{i}_{oqi}} \\ 
        & {{{\dot{i}}}_{odi}}=-\frac{{{R}_{ci}}}{{{L}_{ci}}}{{i}_{odi}}+{{\omega }_{i}}{{i}_{oqi}}+\frac{1}{{{L}_{ci}}}{{v}_{odi}}-\frac{1}{{{L}_{ci}}}{{v}_{bdi}} \\ 
        & {{{\dot{i}}}_{oqi}}=-\frac{{{R}_{ci}}}{{{L}_{ci}}}{{i}_{oqi}}-{{\omega }_{i}}{{i}_{odi}}+\frac{1}{{{L}_{ci}}}{{v}_{oqi}}-\frac{1}{{{L}_{ci}}}{{v}_{bqi}} 
    \end{aligned}\right.\label{eqn_MG_model_last}
\end{align}
where $i_{ldi},i_{lqi}$ denote currents at the LC filter inductance; $v_{bdi},v_{bqi}$ denote the voltages at the connection bus in Figure~\ref{fig_DG}.

\subsection*{Parameters}
\begin{table}[H]
\centering
\caption{Parameters of the tested 4-bus MG system ($T_s^{mpc}=0.05$s,\quad$T_s=0.01$s).}\label{table_para_MG1}
    \begin{tabular}{lcccc}
        \hline
        \headrow
& & \thead{DG1} & \thead{DG2} &    \thead{DG3\& DG4} \\
\hline
\hiderowcolors
        \multirow{11}{*}{DGs}   &$m_P$      &$6.28\times10^{-5}$        &$9.42\times10^{-5}$    &$12.56\times10^{-5}$ \\
                                &$n_{Q}$    &$0.5\times10^{-3}$         &$0.75\times10^{-3}$    &$1\times10^{-3}$ \\
                                &$R_{f}$    &0.1 $\Omega$               &0.1 $\Omega$           &0.1 $\Omega$ \\
                                &$L_{f}$    &1.35 mH                    &1.35 mH                &1.35 mH \\
                                &$C_{f}$    &47$ \mu$F                  &47$ \mu$F              &47 $\mu$F \\
                                &$R_{c}$    &0.02 $\Omega$              &0.02 $\Omega$          &0.02 $\Omega$ \\
                                &$L_{c}$    &2 mH                       &2 mH                   &2 mH \\
                                &$K_{Pv}$   &0.05                       &0.05                   &0.1 \\
                                &$K_{Iv}$   &390                        &390                    &420 \\
                                &$K_{Pc}$   &10.5                       &10.5                   &15 \\
                                &$K_{Ic}$   &$1.6\times10^{4}$          &$1.6\times10^{4}$      &$2\times10^{4}$ \\ 
        \hline
        \multirow{3}{*}{Lines}  &Line1   &\multicolumn{3}{c}{${ R=0.23~\Omega,~L=318~\mu}$H} \\
                                &Line2   &\multicolumn{3}{c}{${ R=0.35~\Omega,~L=1847~\mu }$H} \\
                                &Line3   &\multicolumn{3}{c}{${ R=0.23~\Omega,~L=318~\mu }$H} \\ \hline
        \multirow{4}{*}{RL Loads} &Load1 &\multicolumn{3}{c}{${ R=2~\Omega,~L=6.4~}$mH} \\                                                                             
                                &Load2   &\multicolumn{3}{c}{${ R=4~\Omega,~L=9.6~}$mH} \\
                                &Load3   &\multicolumn{3}{c}{${ R=6~\Omega,~L=12.8~}$mH} \\
                                &Load4   &\multicolumn{3}{c}{${ R=6~\Omega,~L=12.8~}$mH} \\
        \hline
        \multirow{3}{*}{Control Parameters} & DMPC  &\multicolumn{3}{c}{$v_{ref}=311 (220\sqrt{2})$, $H=10$} \\
        & Thresholds & \multicolumn{3}{c}{$e_{opt}=0.1,{e}_{com}=0.1$}\\
        & Observer & \multicolumn{3}{c}{$\varpi=2.5,\left[\omega_{0},\omega_{1},\omega_{2}\right]=\left[1,2,3\right]$} \\ 
        \hline
    \end{tabular}
\end{table}




\bibliography{references}



\end{document}